\documentclass[aip,jcp,amsmath,amssymb,floatfix,citeautoscript,reprint]{revtex4-1}

\usepackage{graphicx}
\usepackage{dcolumn}
\usepackage{bm}
\usepackage{subcaption}

\usepackage{braket}
\usepackage[usenames,dvipsnames]{color}
\usepackage[version=3]{mhchem}
\usepackage{ragged2e}
\usepackage[labelsep=period, format=plain, justification=justified, font=footnotesize]{caption}
\usepackage{times}
\usepackage{txfonts}
\usepackage{relsize}
\usepackage{tabularx, booktabs, makecell, multirow}
\newcolumntype{L}{>{\raggedright\arraybackslash}X}
\newcolumntype{R}{>{\raggedleft\arraybackslash}X}

\DeclareCaptionJustification{myjust}{\justifying}
\newcommand{\RNum}[1]{\uppercase\expandafter{\romannumeral #1\relax}}

\begin{document}

\title{Characterizing and contrasting structural proton transport mechanisms in azole hydrogen bond networks using \textit{ab initio} molecular dynamics}

\author{Austin O. Atsango}
\affiliation{Department of Chemistry, Stanford University, Stanford, California 94305, USA}

\author{Mark E. Tuckerman}
\email{mark.tuckerman@nyu.edu}
\affiliation{Department of Chemistry, New York University, New York, NY 10003, USA}
\affiliation{Courant Institute of Mathematical Science, New York University, New York, NY 10012, USA}
\affiliation{NYU-ECNU Center for Computational Chemistry at NYU Shanghai, 3663 Zhongshan Road North, Shanghai 200062, China}

\author{Thomas E. Markland}
\email{tmarkland@stanford.edu}
\affiliation{Department of Chemistry, Stanford University, Stanford, California 94305, USA}

\date{\today}

\begin{abstract}
Imidazole and 1,2,3-triazole are promising hydrogen-bonded heterocycles that conduct protons via a structural mechanism and whose derivatives are present in systems ranging from biological proton channels to proton exchange membrane fuel cells. Here, we leverage multiple time-stepping to perform {\it ab initio} molecular dynamics of imidazole and 1,2,3-triazole at the nanosecond timescale. We show that despite the close structural similarities of these compounds, their proton diffusion constants vary by over an order of magnitude. Our simulations reveal the reasons for these differences in diffusion constants, which range from the degree of hydrogen-bonded chain linearity to the effect of the central nitrogen atom in 1,2,3-triazole on proton transport. In particular, we uncover evidence of two ``blocking" mechanisms in 1,2,3-triazole, where covalent and hydrogen bonds formed by the central nitrogen atom limit the mobility of protons. Our simulations thus provide insights into the origins of the experimentally observed 10-fold difference in proton conductivity.
\end{abstract}

{\maketitle}

\normalsize
The structural diffusion mechanism of excess protons in hydrogen-bonded systems, which involves a series of intermolecular proton transfer reactions, is responsible for the high rate of proton diffusion observed in liquids such as water~\cite{tuckerman1995ab,tuckerman1995ab2,agmon1995grotthuss,marx1999nature,vuilleumier1999transport,schmitt1999computer,Zhu2002,tuckerman2002nature,woutersen2006ultrafast,markovitch2008special,berkelbach2009concerted,marx2010aqueous,agmon2016protons,Napoli2018,Yuan2019,Roy2020}, phosphoric acid~\cite{Viliauskas2012}, and imidazole~\cite{Chen2009,long2020elucidating}, in solids such as the superprotonic phases of cesium dihydrogen phosphate~\cite{Boysen2004,Lee2008,Kim2013} and cesium hydrogen sulfate~\cite{Wood2007}, and in ionic solids such as doped ammonium perchlorate~\cite{Rosso2003}. This mechanism is also vital for proton transport in systems ranging from biological proton pumps~\cite{Quan2020} to proton exchange membrane (PEM) fuel cells~\cite{SCHUSTER2008}. Most experimental and theoretical studies of structural proton transport have focused on water~\cite{tuckerman1995ab,tuckerman1995ab2,agmon1995grotthuss,marx1999nature,vuilleumier1999transport,schmitt1999computer,Zhu2002,tuckerman2002nature,woutersen2006ultrafast,markovitch2008special,berkelbach2009concerted,marx2010aqueous,agmon2016protons,Napoli2018,Yuan2019,Roy2020}, which is widely used to assist proton conduction in PEM fuel cells via the perfluorosulfonic polymer Nafion.~\cite{colomban1992proton} However, PEM fuel cells that rely on water-assisted proton transport have a limited operational temperature (up to $\sim$353~K) due to water's low boiling point and suffer from chemical short-circuiting due to the high electro-osmotic drag of water.~\cite{kreuer2004transport} In addition, the influence of morphology and the chemistry of the anionic functional groups have yet to be fully clarified~\cite{Trigg2018,Zelovich2021}.  Given these challenges, there is significant interest in exploring alternatives such as organic heterocycles due to their role in charge transfer in biological systems~\cite{decoursey2003voltage} and their ability to be chemically integrated into PEM materials.~\cite{kreuer2004transport,Schuster2004,kawada1970protonic} A particularly intriguing pair of liquid heterocycles is imidazole and 1,2,3-triazole, both of which are efficient structural proton conductors~\cite{kawada1970protonic,zhou2005promotion} that remain liquid in the temperature ranges 363 - 530~K and 296 - 476~K respectively and, thus, also offer the benefit of a high operational temperature.

\begin{figure}[h!]
\centering
\includegraphics[width=0.45\textwidth]{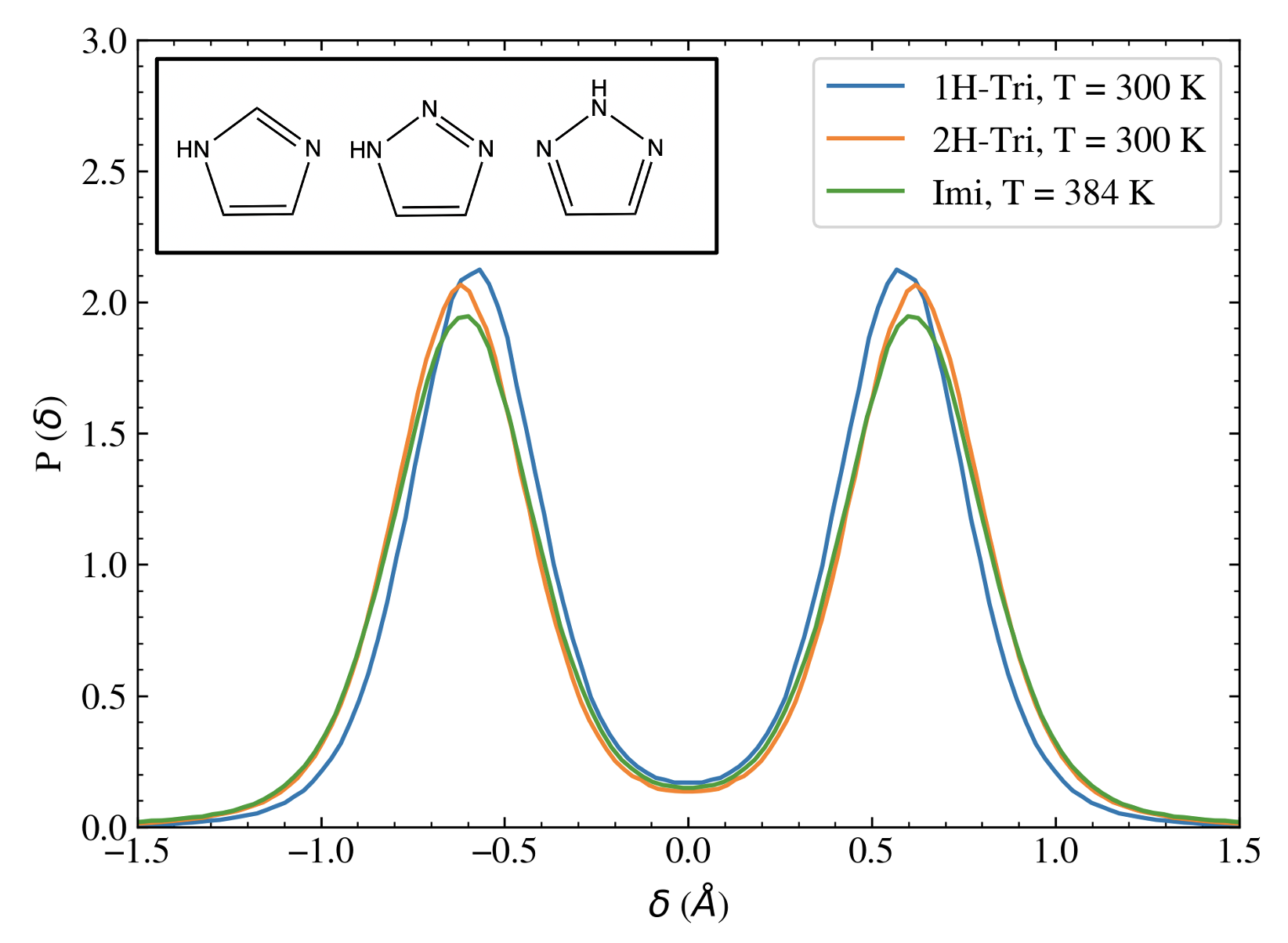}
\caption{Probability distributions along the proton transfer progress coordinate, $\delta$, for H* in Imi, 1H-Tri, and 2H-Tri. The inset shows molecular structures of Imi, 1H-Tri, and 2H-Tri from left to right.}
\label{fig:delta_plot}
\end{figure}

Due to their geometry, imidazole and 1,2,3-triazole form low-dimensional hydrogen-bonded structures, specifically chain structures, that differ significantly from the 3-dimensional networks observed in water. In a recent study, we highlighted the importance of hydrogen-bonded chains in the structural proton transport mechanism of imidazole by showing how one can identify three distinct regimes of proton transport corresponding to short-time exchange of protons between pairs of molecules, intermediate-time exploration of the proton along a particular hydrogen-bonded chain, and long-time chain rearrangement.~\cite{long2020elucidating} 1,2,3-triazole, while almost identical in structure to imidazole, contains an extra nitrogen atom that can accept an additional hydrogen bond, allowing it potentially to form  a structurally different H bond network. This seemingly subtle change causes a significant difference in the proton conductivity to that of imidazole, with the conductivity of imidazole exceeding that of 1,2,3-triazole by an order of magnitude when both liquids are just above their melting points (363 K and 296 K respectively).~\cite{zhou2005promotion,kawada1970protonic}
The extra nitrogen atom also allows 1,2,3-triazole to exist as a mixture of two tautomers, 1-{\it H}-1,2,3-triazole and 2-{\it H}-1,2,3-triazole, which differ in the position of the nitrogen that forms a covalent bond with H (inset in Fig.~\ref{fig:delta_plot}). 2-{\it H}-1,2,3-triazole is the dominant tautomer in the gas phase\cite{Begtrup1988}, while the solid consists of a 1:1 mixture of both tautomers\cite{Goddard1997}. In the liquid phase, the identity of the dominant tautomer is still a matter of debate, although a recent combined experimental and simulation study has suggested that 2-$\textit{H}$-1,2,3-triazole may be the dominant tautomer.~\cite{bellagamba2013tautomerism} Contrasting the hydrogen bonding and proton transport properties of these two triazole tautomers with imidazole thus offers the opportunity to elucidate the subtle interplay of this additional hydrogen bonding interaction with the length and topology of the hydrogen bond chains formed and the dynamics arising from them.

In this study, we employ the r-RESPA multiple time stepping (MTS) scheme~\cite{tuckerman1992reversible,luehr2014multiple,marsalek2017quantum} to perform nanosecond \textit{ab initio} molecular dynamics (AIMD) simulations of an excess proton in liquid imidazole and in liquids of each of the two tautomers of 1,2,3-triazole just above their respective melting points. These MTS-accelerated AIMD simulations allow us to capture the bond making and breaking involved in the proton transport process. Our simulations contain 577 and 513 atoms for imidazole and both tautomers of 1,2,3-triazole respectively, and thus are able to capture the hydrogen-bonded chains along which proton transport occurs. By performing dynamics in excess of a nanosecond for each liquid, we are able to access long time scale hydrogen bond and proton transport processes. We perform separate simulations of 1-$\textit{H}$-1,2,3-triazole and 2-$\textit{H}$-1,2,3-triazole in order to establish how the topologically different hydrogen bond networks formed by each tautomer affect proton transport. Based on these results, we demonstrate the crucial role of the covalent and hydrogen bonds formed by the central nitrogen atom in 1,2,3-triazole in determining the dynamics of the hydrogen bonds and proton transport in these liquids.

Table~\ref{table:msds} shows the molecular and proton diffusion coefficients obtained from our AIMD simulations for imidazole (Imi) at 384~K and 1-\textit{H}-1,2,3-triazole (1H-Tri) and 2-\textit{H}-1,2,3-triazole (2H-Tri) at 300~K; these temperatures were chosen as they are just above the compounds' respective melting points. Despite the similarity of their chemical structures, the proton diffusion coefficients of 1H-Tri and 2H-Tri tautomers are lower than that of Imi by factors of $6$ and $25$, respectively, a range that spans the factor of 10 ratio in the conductivity of imidazole and 1,2,3-triazole observed experimentally\cite{zhou2005promotion,kawada1970protonic}. However, even though 1H-Tri has a smaller observed proton diffusion coefficient than Imi, it has a similarly large enhancement of its proton diffusion coefficient (D$_p$) over that of the molecule itself (D$_{mol}$), i.e., 8.9 compared to 8.3 for Imi. This indicates that for both of these molecules, proton diffusion is dominated by a structural (Grotthuss-type) mechanism that allows for highly efficient and selective proton transport. In contrast, 2H-Tri has a molecular diffusion coefficient 2.2 fold higher than that of 1H-Tri, but the fact that the 2H-Tri ratio D$_p$/D$_{mol} \sim 1$ indicates that structural diffusion contributes negligibly to proton motion in this system. SI Table~\ref{table:decomposition} shows the decomposition of D$_p$, i.e. the diffusion of the protonated species, into its vehicular and structural components.

\begin{table}[ht]
\setcellgapes{3pt}
\makegapedcells
\centering
\begin{tabularx}{\linewidth}{LRRR}
\toprule[1.5pt]
System & Imidazole (384~K) & 1-\textit{H}-1,2,3-triazole (300~K) & 2-\textit{H}-1,2,3-triazole (300~K)\\
\hline
D$_p$ & 0.47 $\pm$ 0.04 & 0.08 $\pm$ 0.02 & 0.020 $\pm$ 0.003\\
D$_{mol}$ & 0.0663 $\pm$ 0.0004 & 0.009 $\pm$ 0.001 & 0.019 $\pm$ 0.002\\
\bottomrule[1.5pt]
\end{tabularx}
\caption{Diffusion coefficients in $\AA^2$/ps for Imi, 1H-Tri, and 2H-Tri. D$_p$ denotes the proton diffusion coefficient, while D$_{mol}$ denotes the molecular diffusion coefficient.}
\label{table:msds}
\end{table}

\begin{figure}[h!]
    \centering
    \includegraphics[width=0.45\textwidth]{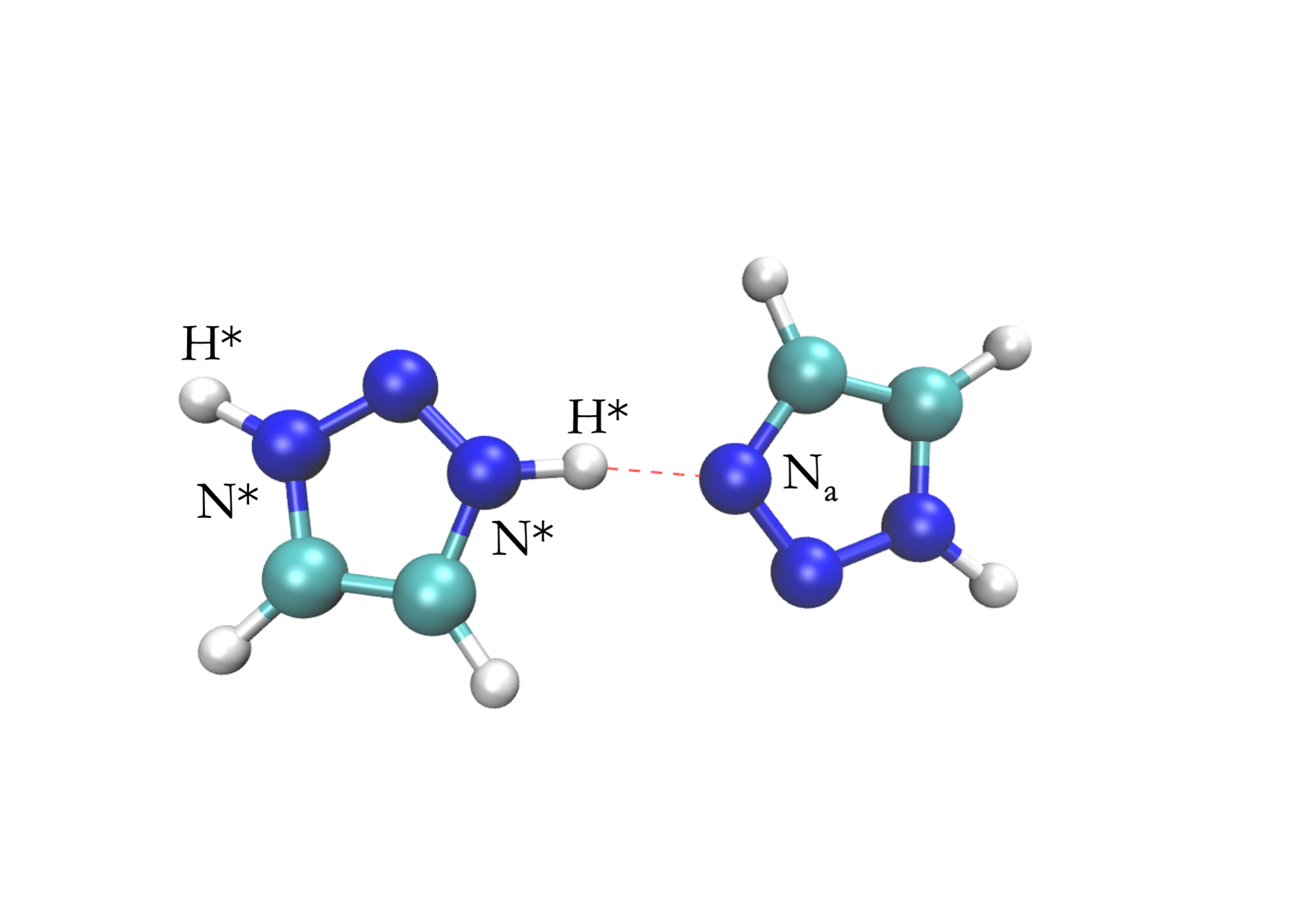}
    \caption{Hydrogen bond pair featuring a protonated molecule (left) and an unprotonated molecule (right) in 1H-Tri. Positions of the atoms labelled $\mathrm{N^*}$, $\mathrm{H^*}$, and $\mathrm{N_a}$ are used to compute $\delta=r_\mathrm{N_aH^*}-r_\mathrm{N^*H^*}$.}
    \label{fig:H-bond-pair}
\end{figure}

In order to understand the origins of observed differences in the proton diffusion coefficients obtained from our AIMD simulations, we begin by comparing the extent of sharing of the excess proton in Imi, 1H-Tri, and 2H-Tri. To this end, we use the proton transfer progress coordinate, $\delta=r_\mathrm{N_aH^*}-r_\mathrm{N^*H^*}$, where $\mathrm{H^*}$ and $\mathrm{N^*}$ are the excess proton and its covalently bonded nitrogen atom respectively, and $\mathrm{N_a}$ is the acceptor nitrogen atom to which $\mathrm{H^*}$ is hydrogen-bonded (Fig.~\ref{fig:H-bond-pair}). $\delta$ measures how much the excess proton deviates from a position that is equidistant between $\mathrm{N^*}$ and $\mathrm{N_a}$. Thus $\delta=0$ is a necessary (but not always sufficient) condition for a proton transfer event to occur. Figure~\ref{fig:delta_plot}, which shows the probability distribution along $\delta$, reveals that despite the 1H-Tri and 2H-Tri simulations having been performed at a temperature 84 K lower than Imi, all three systems have almost identical $\delta$ probability distributions, especially around $\delta  = 0$ where the probability is approximately 10-fold lower than that at the location of maximum probability ($\delta=\pm 0.6~\AA$) in all three systems. The free energy barriers to proton transfer for the three systems at their respective temperatures are also similar: 1.6 kcal/mol, 1.6 kcal/mol, and 1.5 kcal/mol for Imi, 2H-Tri, and 1H-Tri respectively. The similarity in probability distributions demonstrates that there is very little difference in the ease with which protons can hop between pairs of molecules, a fact that is corroborated by the similar hydrogen bond strengths observed across the three systems (see SI Section~\ref{sec:bond_overview}). This is also consistent with the similar values obtained for the short time decay of the population correlation function of protonated species~\cite{Chandra2007,Tuckerman2010}: 0.21~ps, 0.30~ps, and 0.25~ps for Imi, 1H-Tri, and 2H-Tri respectively (see SI Fig.~\ref{fig:corr_funs}). This timescale is associated with ``rattling" of the proton, defined as transient hops of the proton that occur between neighboring molecules and are reversed by the next proton hop. In all three systems, over 90\% of all observed proton hops are rattling events(see SI Table~\ref{table:hopstat}), and the percentages of productive proton hops (i.e. those that aren't reversed by the next proton hop) are 8.0\%, 2.6\%, and 1.5\% respectively for Imi, 1H-Tri, and 2H-Tri, which matches the ordering of measured proton diffusion coefficients (D$_p$(Imi) $>$ D$_p$(1H-Tri) $>$ D$_p$(2H-Tri)). As such, elucidating the differences in the diffusion coefficients requires consideration of the mechanisms that allow for longer-range proton motion.

\begin{figure}[h]
    \centering
    \begin{subfigure}{0.5\textwidth}
    \includegraphics[width=0.9\textwidth]{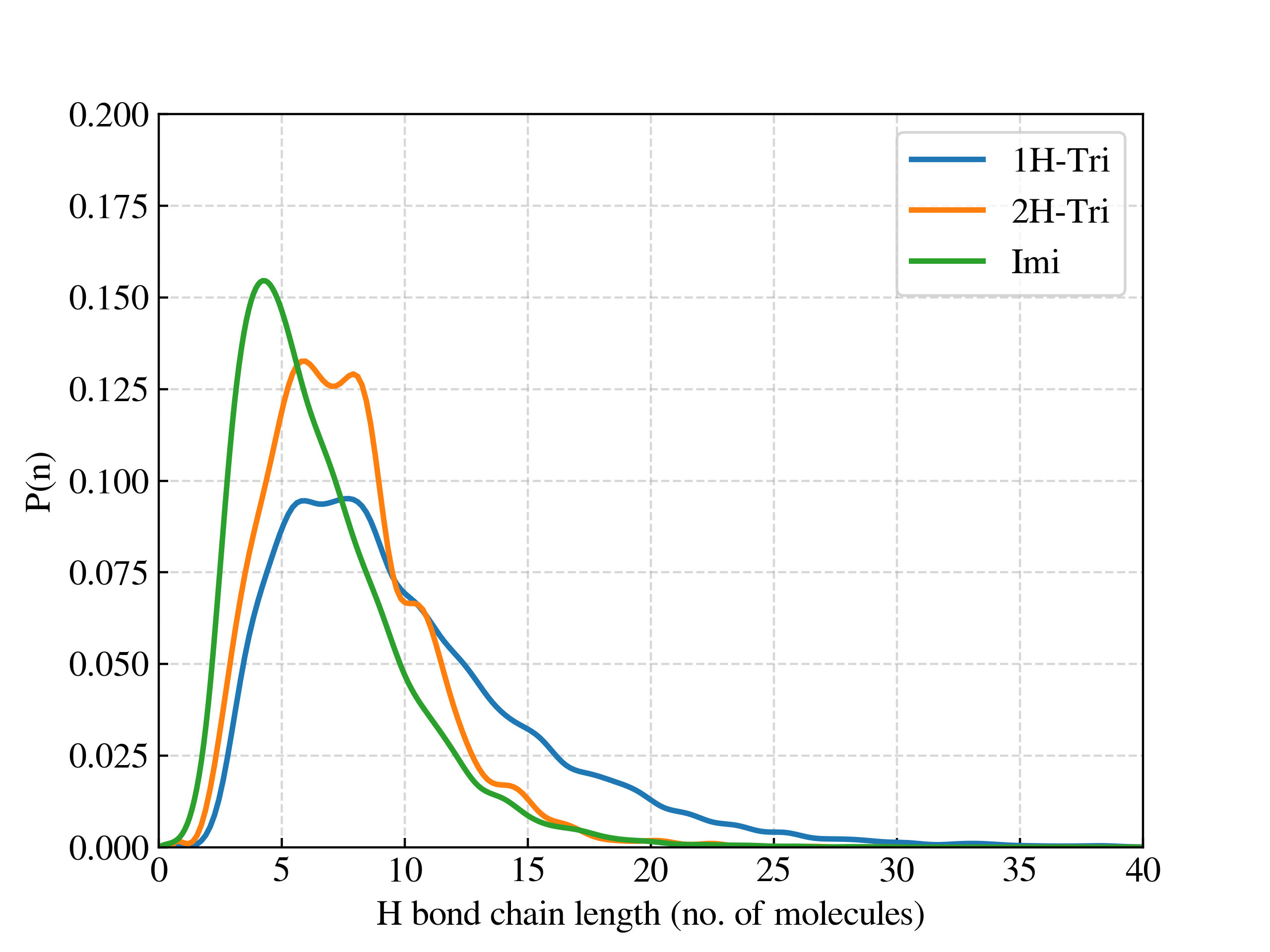}
    \caption{Probability distributions for the lengths of hydrogen bond chains around the protonated molecule in Imi, 1H-Tri, and 2H-Tri}
    \label{fig:chain_length_dist}
    \end{subfigure}
    \begin{subfigure}{0.5\textwidth}
    \centering
    \includegraphics[width=0.9\textwidth]{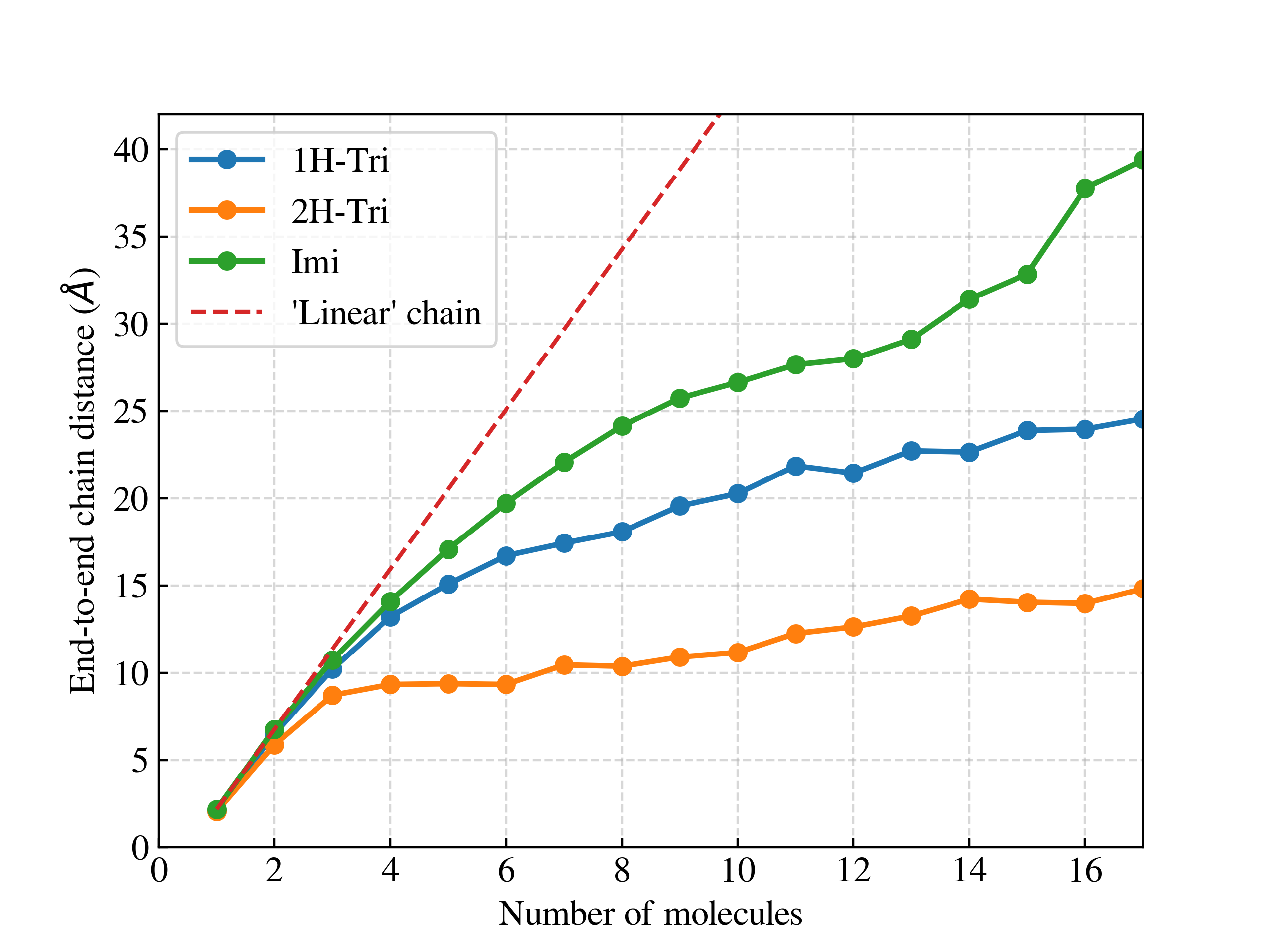}
    \caption{Plot showing how the end-to-end distance across a hydrogen bond chain centered around the protonated molecule varies with the number of molecules present in the chain for Imi, 1H-Tri, and 2H-Tri}
    \label{fig:chain_linearity}
    \end{subfigure}
    \label{fig:chains}
    \caption{}
\end{figure}

Imidazole forms long pseudo-one-dimensional hydrogen bond chains in the liquid phase that play a vital role in its ability to transport protons.~\cite{Mnch2001,long2020elucidating} Indeed, we have recently shown that in imidazole, the protonated species can reach an average steady-state mean square displacement (MSD) of 40~$\AA^{2}$ from its initial position without leaving an intact hydrogen-bonded imidazole chain. However, while an Imi molecule only has two nitrogen atoms through which it can accept one hydrogen bond and donate one, 1,2,3-triazole contains an additional nitrogen atom, which allows the molecule to accept up to two hydrogen bonds and donate one. To assess how this changes the length and topology of the hydrogen bond chains in triazole tautomers, Fig.~\ref{fig:chain_length_dist} shows the probability distribution of the number of molecules that make up the hydrogen-bonded chain formed around the protonated molecule in the three liquids. The figure demonstrates that 1H-Tri and 2H-Tri both form longer chains on average than Imi with means of 8.6, 7.1, and 6.6 molecules respectively. This ordering does not match that of observed diffusion coefficients. For example, 2H-Tri has the lowest proton diffusion coefficient with no enhancement from a structural diffusion mechanism, and yet its average hydrogen bond chain length lies between that of Imi and 1H-Tri. One possible reason for this is that the motifs formed by the triazoles can lead to chains that contain more kinks and thus have significantly shorter end-to-end distances. Fig.~\ref{fig:chain_linearity} shows the average end-to-end distances for 1H-Tri, 2H-Tri, and Imi for chains consisting of different numbers of molecules centered around the molecule holding the proton defect. From this, one can see that, indeed, as the number of molecules in the chain increases, the end-to-end distance of the 2H-Tri chains quickly deviates from the behavior expected for purely linear chains, and for more than 3 or 4 molecules in the chain, the end-to-end distance saturates, indicating the presence of highly non-linear chains containing significant numbers of kinks and/or loops. This is in stark contrast with Imi, where even at chain lengths of 16 molecules, the end-to-end distance still increases markedly with each molecule in the chain (see Fig.~\ref{fig:chain_linearity}). 1H-Tri shows an intermediate behavior between 2H-Tri and Imi. Slower proton diffusion in 2H-Tri is thus at least partially caused by non-linear chains that pack inefficiently and are similar to what is observed in pyrazole, which must form hydrogen bonds from the 1 and 2 positions, and is a poor proton conductor.~\cite{nagamani2010proton}

\begin{figure*}
    \centering
    \includegraphics[width=0.8\textwidth]{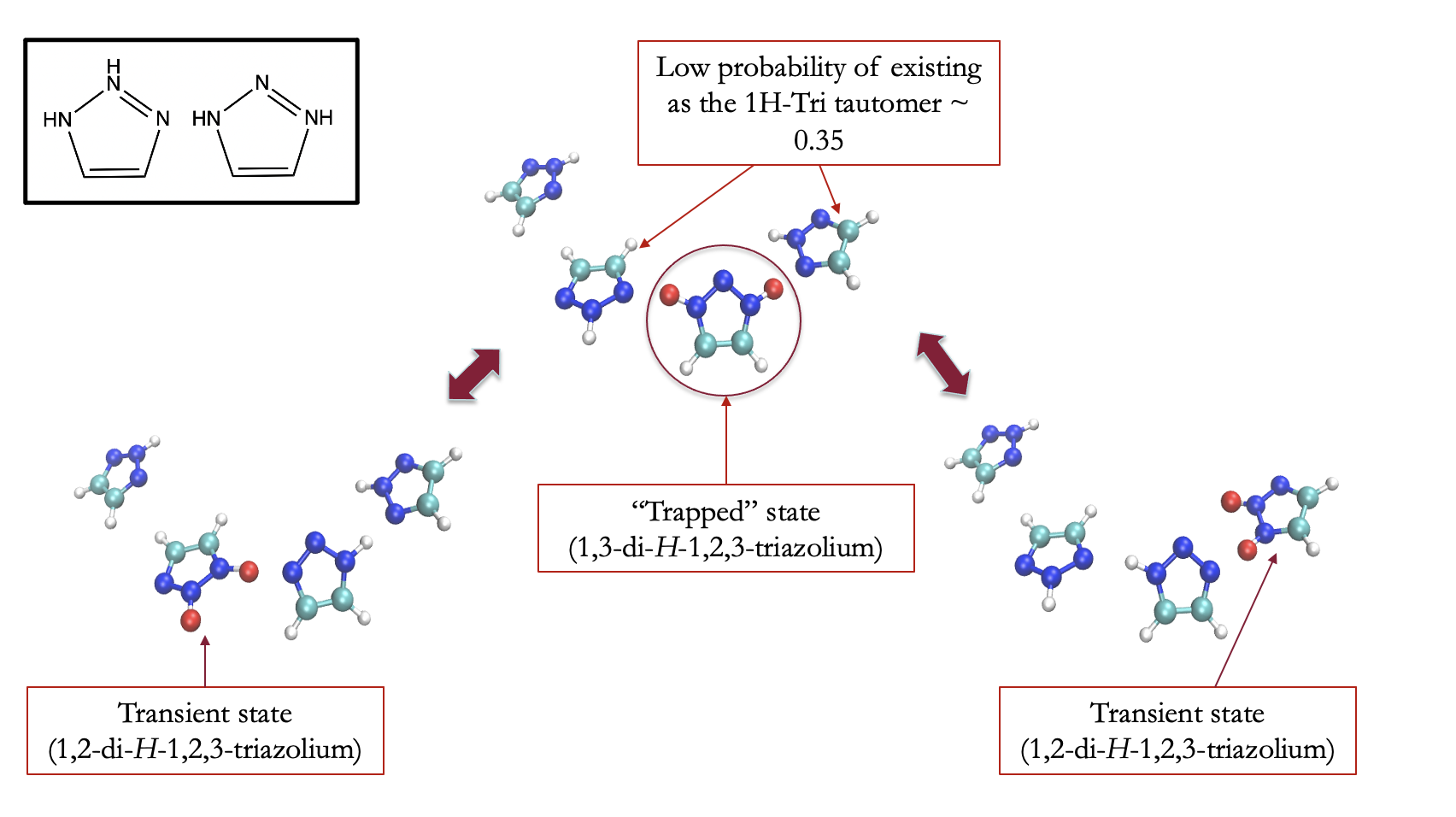}
    \caption{Demonstration of the excess proton ``trap". In the central image, molecules in the first solvation shell of 1,3-di-\textit{H}-1,2,3-triazolium have a high probability ($\sim$ 0.65) of existing as 2H-Tri. As such, the movement of an excess proton to either side of the triazolium molecule leads to the formation of the less favored 1,2-di-\textit{H}-1,2,3-triazolium, thus impeding proton transport through the network. Inset shows molecular structures of 1,2-di-\textit{H}-1,2,3-triazolium and 1,3-di-\textit{H}-1,2,3-triazolium respectively.}
    \label{fig:proton_trap}
\end{figure*}

While the non-linearity of hydrogen bond chains in 2H-Tri partially accounts for its low proton diffusivity, it does not explain why structural diffusion is so slow as to be effectively absent in this system. To explore why, we examine the nature of the protonated form, 1,2,3-triazolium. Much like 1,2,3-triazole, 1,2,3-triazolium has two tautomers: 1,2-di-\textit{H}-1,2,3-triazolium, which has protons at the N1 and N2 positions, and 1,3-di-\textit{H}-1,2,3-triazolium, which has the protons at the N1 and N3 (Fig.~\ref{fig:proton_trap} inset) positions. In our simulations of the 2H-Tri system, although protonation of 2H-Tri forms 1,2-di-$H$-1,2,3-triazolium, the 1,3-di-\textit{H}-1,2,3-triazolium species can form via an intermolecular mechanism wherein a 1,2-di-\textit{H}-1,2,3-triazolium molecule transfers the proton at its N2 position to a neighboring 2H-Tri molecule. This transfer leaves behind a 1H-Tri molecule that can subsequently accept an extra proton at its N3 position, producing the 1,3-di-\textit{H}-1,2,3-triazolium cation. Our analysis of the 2H-Tri trajectory shows that although  1,2-di-\textit{H}-1,2,3-triazolium is initially present at $t=0$, the 1,3-di-\textit{H}-1,2,3-triazolium tautomer dominates, with 94\% of all frames exhibiting this form. This preference creates a trap for the excess proton because the chance of a proton hop to a 2H-Tri molecule, with a proton at the N2 position, is low, as such an event would form the unfavorable, transient 1,2-di-\textit{H}-1,2,3-triazolium. Productive proton hops would thus require the adjacent 2H-Tri molecules to have tautomerized to 1H-Tri. Our simulations show that the probability of observing the 1H-Tri tautomer in molecules adjacent to triazolium is low ($\sim$ 0.35). This probability drops even further to 0.09 one molecule away from the excess proton and stabilizes at $\sim$0.05 at four molecules away. This restricts proton transfer since in most cases, proton hops to adjacent molecules form the less favored 1,2-di-\textit{H}-1,2,3-triazolium (Fig.~\ref{fig:proton_trap}). The high preference for 1,3-di-\textit{H}-1,2,3-triazolium, combined with the low probability of tautomerization of 2H-Tri molecules at and beyond the first solvation shell of the excess proton, are thus factors causing the extremely slow rate of structural proton transport in 2H-Tri.

\begin{figure}[h!]
    \centering
    \includegraphics[width=0.45\textwidth]{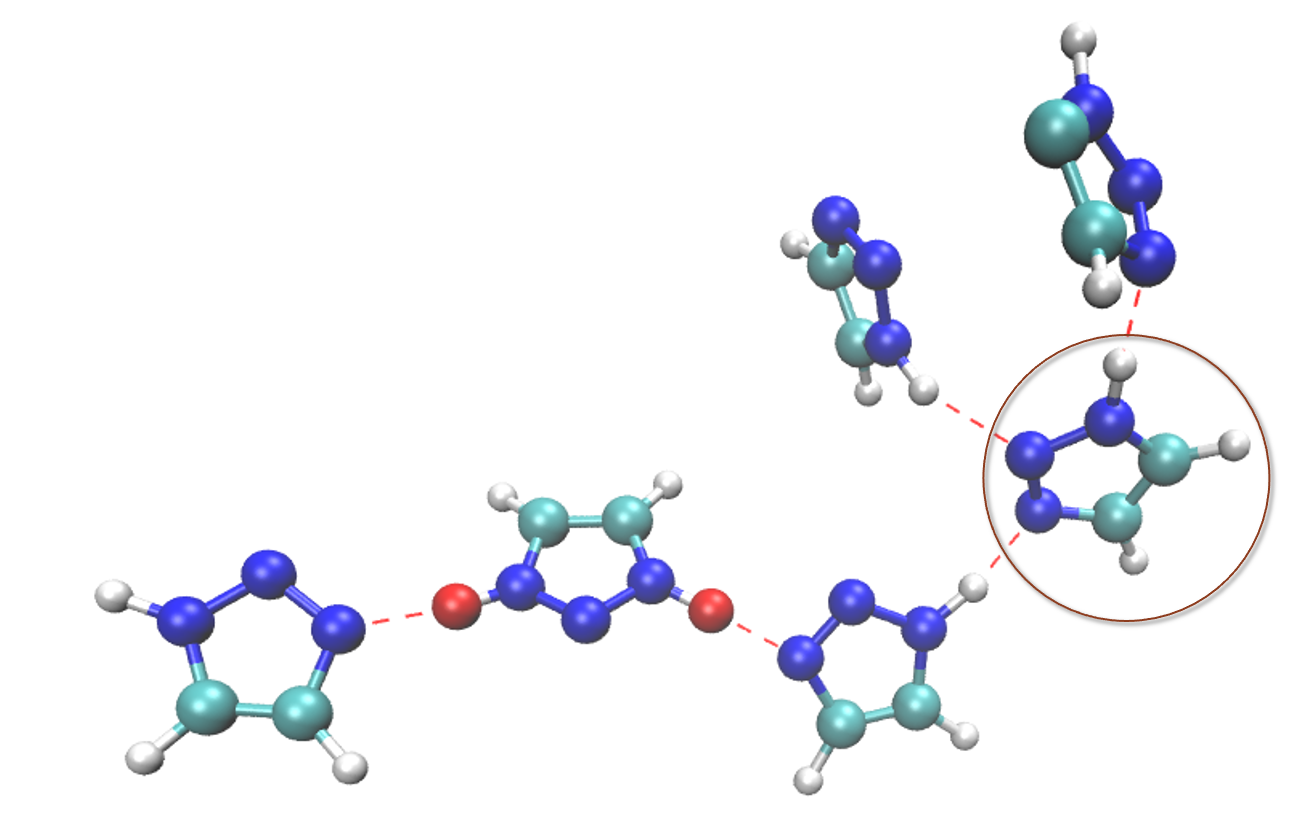}
    \caption{"Blocking" effect caused by receiving a hydrogen bond at the N2 position in 1H-Tri. In this chain consisting of donor hydrogen bonds around the protonated molecule, the circled molecule is unavailable to receive the excess proton because it bears an accepting hydrogen bond at the N2 position. This limits the length of the hydrogen bond chain available for traversal by the excess proton.}
    \label{fig:blocker}
\end{figure}

\begin{figure}[h!]
\centering
\includegraphics[width=0.45\textwidth]{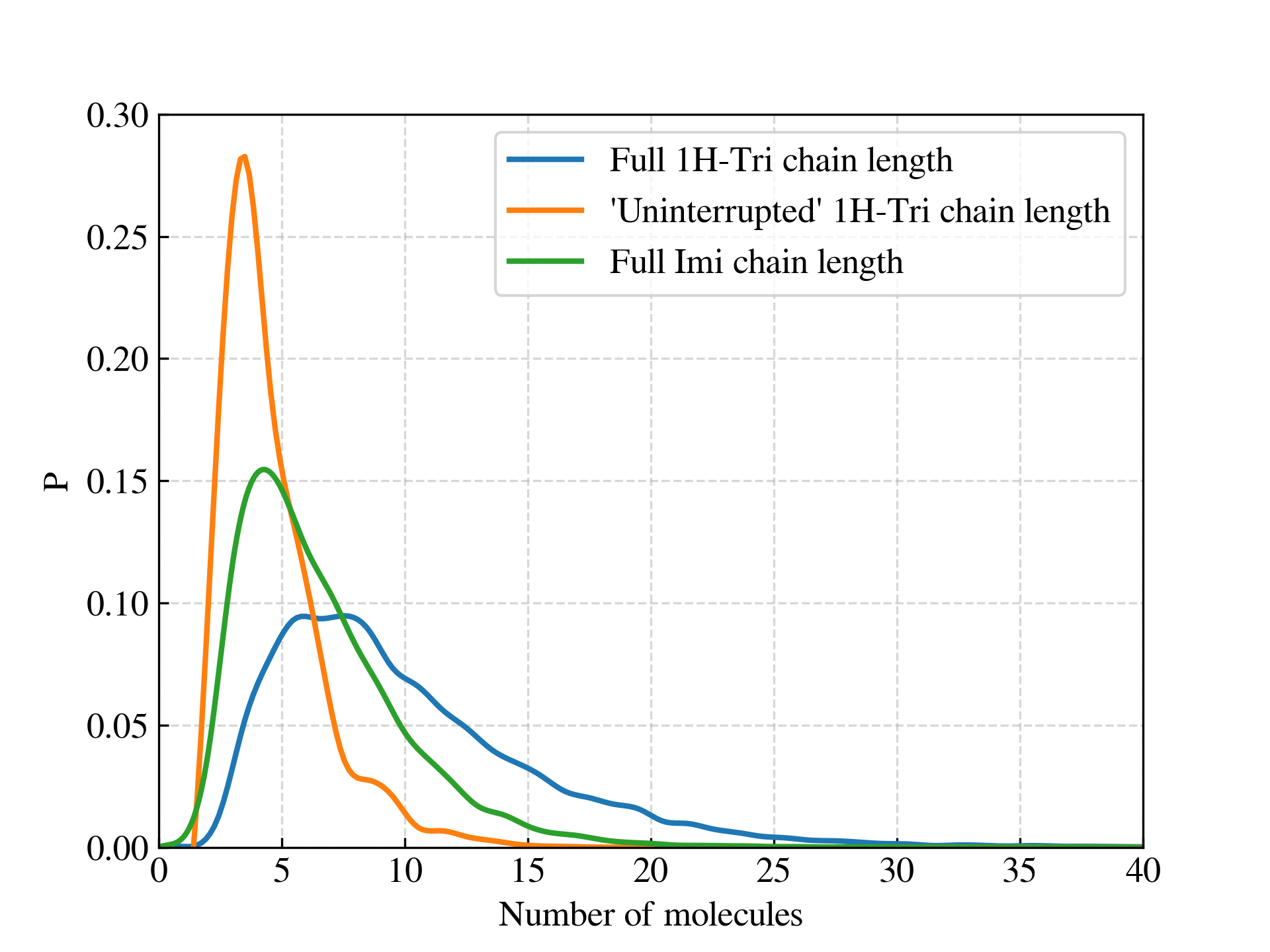}
\caption{Probability distributions for the lengths of hydrogen bond chains around the protonated molecule in Imi and 1H-Tri. The distributions for 1H-Tri include an additional "uninterrupted" case where the hydrogen bond chains are terminated when a molecule either accepts or donates a hydrogen bond at the N2 position.}
\label{fig:chain_lengths_mod}
\end{figure}

Having observed that in 2H-Tri, the covalent bond to hydrogen formed at the N2 position leads to proton trapping and hence the low rate of structural proton diffusion, we now investigate the reason for the difference in the rates of proton transport in Imi and 1H-Tri. In both of these systems, the structural enhancement of proton diffusion over the liquid molecules is $\sim$8, and both form long hydrogen bond chains (Fig.~\ref{fig:chain_length_dist}) with the chains in 1H-Tri being, on average, slightly longer (8.6 molecules vs 6.6 molecules). However, 1H-Tri has a proton diffusion coefficient that is $\sim$6 times lower than imidazole in our simulations when both are just above their melting points (300~K and 384~K, respectively). This seemingly puzzling observation that the longer proton transfer pathways provided by the hydrogen bond chains in 1H-Tri do not lead to faster proton transport can be explained by examining solvation patterns. Due to the additional nitrogen atom (N2 position), 1H-Tri can accept a hydrogen bond at that position. In our simulations, 1H-Tri molecules form a hydrogen bond at the N2 position 16\% of the time. However, for the protonated 1H-Tri molecule, this hydrogen bond is intact only 0.15\% of the time {\it i.e.}, this interaction is disfavored in the protonated species by $\sim$100 fold ($\sim$2.75 kcal/mol) relative to the unprotonated form. This destabilizing hydrogen bond formed from the N2 position suggests a ``blocking" mechanism at play at this N2 position. Specifically, since a 1H-Tri molecule with a hydrogen bond in this position is not solvated in a way that allows it to accommodate the excess proton, it is required that the 1H-Tri molecule not accept a hydrogen bond at the N2 position before it can receive the excess proton. Therefore, 1H-Tri molecules that simultaneously make up the hydrogen bond chain containing the excess proton and accept a hydrogen bond at their N2 position (see Fig.~\ref{fig:blocker}) are not available for proton transfer and act to block structural diffusion pathways. We can incorporate this in our measurement of the length of hydrogen bond chains in 1H-Tri by defining a new ``uninterrupted" chain that terminates every time it encounters a molecule that accepts a hydrogen bond at its N2 position. The new criterion also excludes molecules that donate hydrogen bonds at the N2 position, {\it i.e.}, molecules that have tautomerized to 2H-Tri, since the probability of forming the corresponding 1,2-di-{\it H}-1,2,3-triazolium in the 1H-Tri system is incredibly low ($2\times 10^{-4}$). The results of this analysis are shown in Fig.$~\ref{fig:chain_lengths_mod}$, which contains both the original and uninterrupted hydrogen-bonded chain lengths. Here, we observe that the average uninterrupted chain length in 1H-Tri (4.2) is now shorter than that of imidazole (6.6), which is in line with the trend in proton diffusion coefficients. These results demonstrate that proton transport is slower in 1H-Tri than imidazole because 1H-Tri has shorter uninterrupted chain lengths and hence shorter ``effective" proton transfer pathways.

In order to provide further verification of the length of the hydrogen-bonded chain along which a proton can diffuse before reaching an N2 hydrogen bond acceptor site, we compute the length of hydrogen bond chain available to the excess proton in 1H-Tri using a random walk discrete-time Markov chain model for intra-chain proton transport. This model allows us to predict where the MSD of the proton would saturate if diffusion was limited to a one-dimensional chain consisting of $n$ molecules spaced apart by a distance $d$. The steady-state mean squared displacement, $\mathrm{MSD}(\infty)$, is given by the formula:

\begin{equation}
\mathrm{MSD}(\infty) = \frac{n^2+2}{6} d^2
\end{equation}

Substituting the average uninterrupted chain length ($n=4.2$) and using the most probable N*$-$N hydrogen bond length as the inter-site distance ($d=2.7 \AA$) in 1H-Tri (SI Section~\ref{sec:bond_overview}), we obtain a steady-state MSD of 14 $\AA^2$, which is in good agreement with the MSD at the onset of the linear regime for 1H-Tri of $\sim$20 $\AA^2$ (SI Section~\ref{sec:msd_plots}). The change in the slope of the MSD at that distance reflects the transition from diffusion on a single chain (limited to MSD($\infty$)) to the long-range diffusion that requires chain rearrangement. 

\begin{figure*}
    \centering
    \begin{minipage}{0.3\textwidth}
       \centering
       \includegraphics[width=\textwidth]{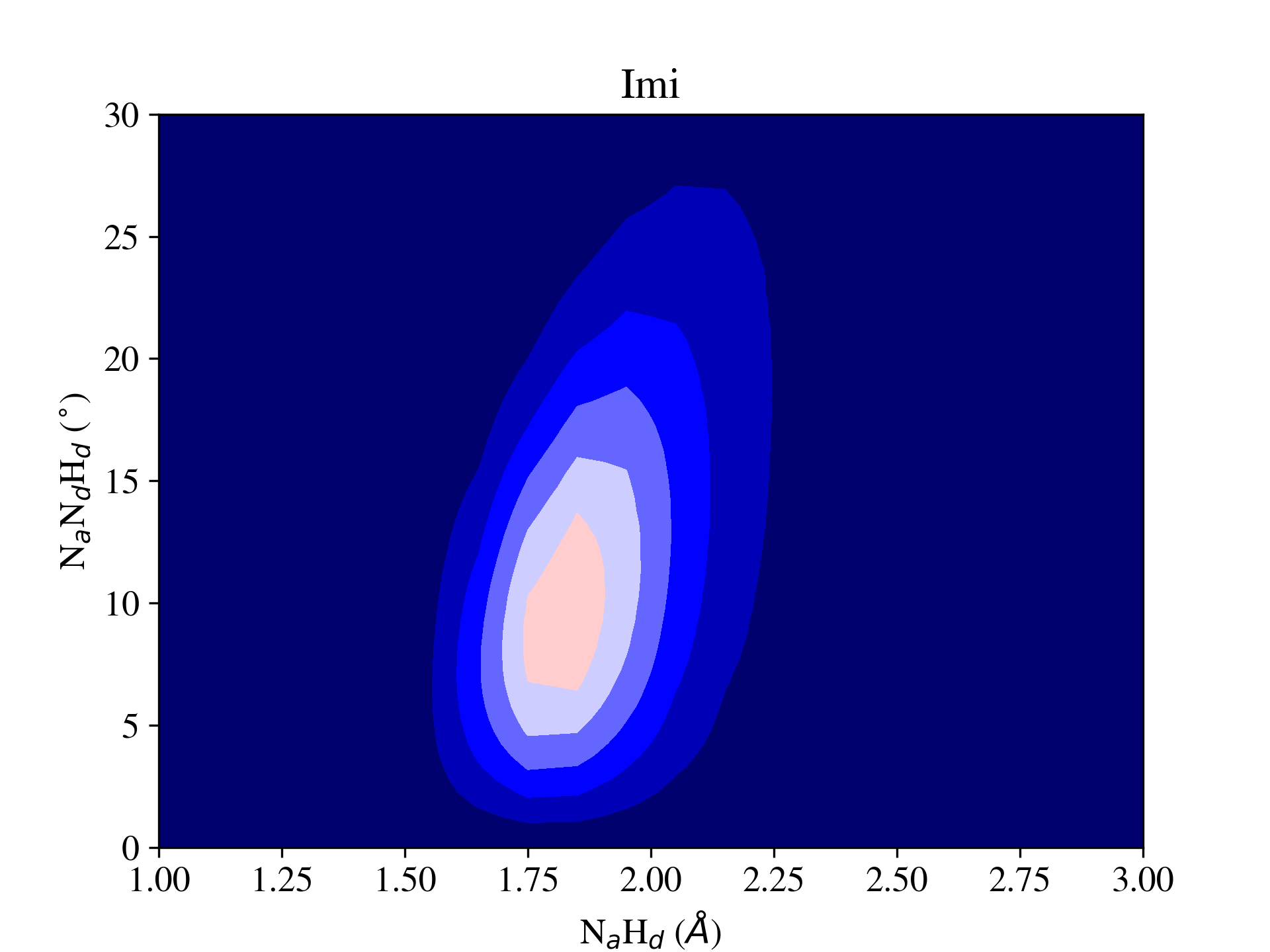}
   \end{minipage}
   \begin{minipage}{0.3\textwidth}
      \includegraphics[width=\textwidth]{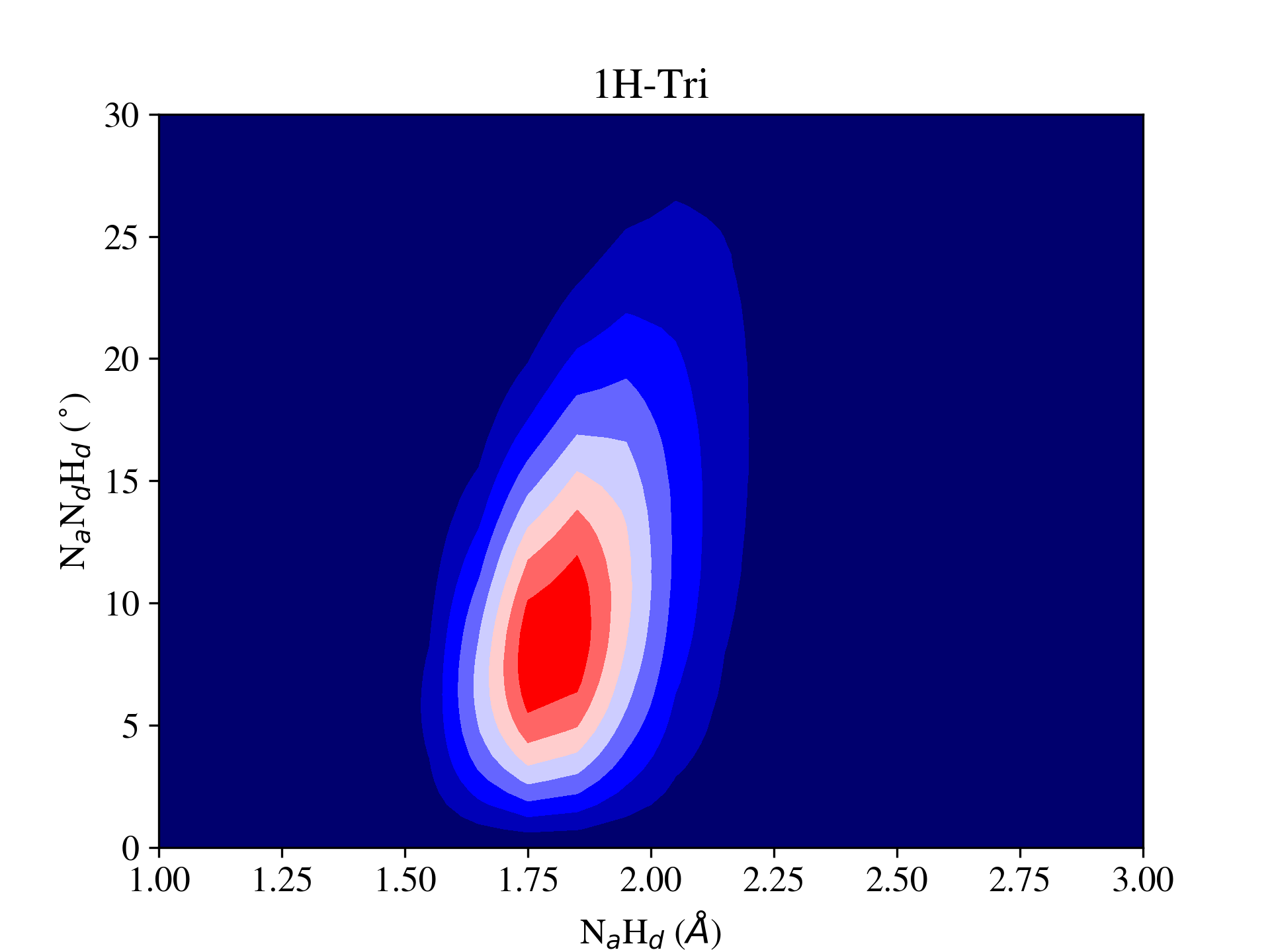}
   \end{minipage}
   \begin{minipage}{0.3\textwidth}
       \centering
       \includegraphics[width=\textwidth]{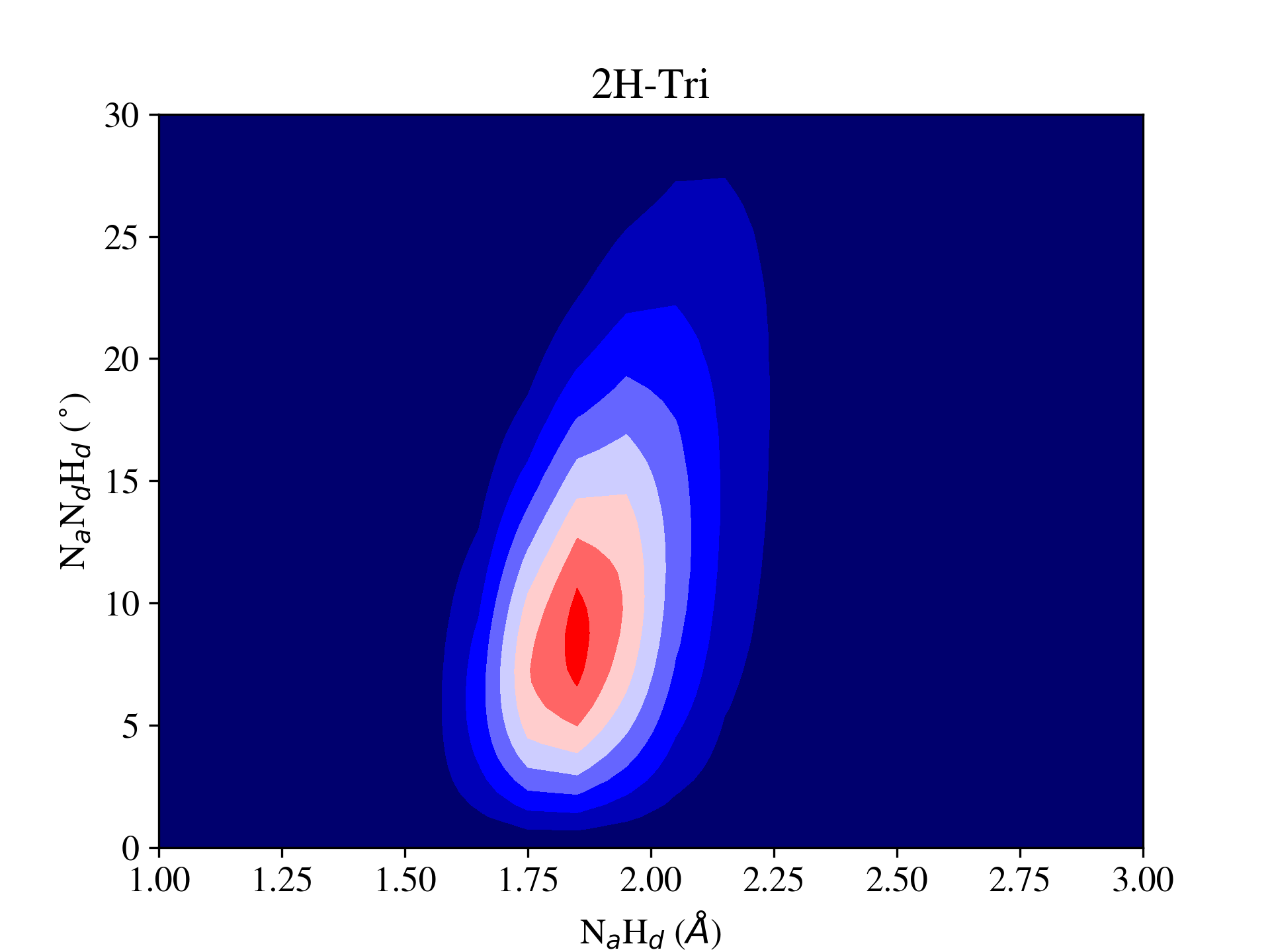}
   \end{minipage}
   \begin{minipage}{0.3\textwidth}
       \centering
       \includegraphics[width=\textwidth]{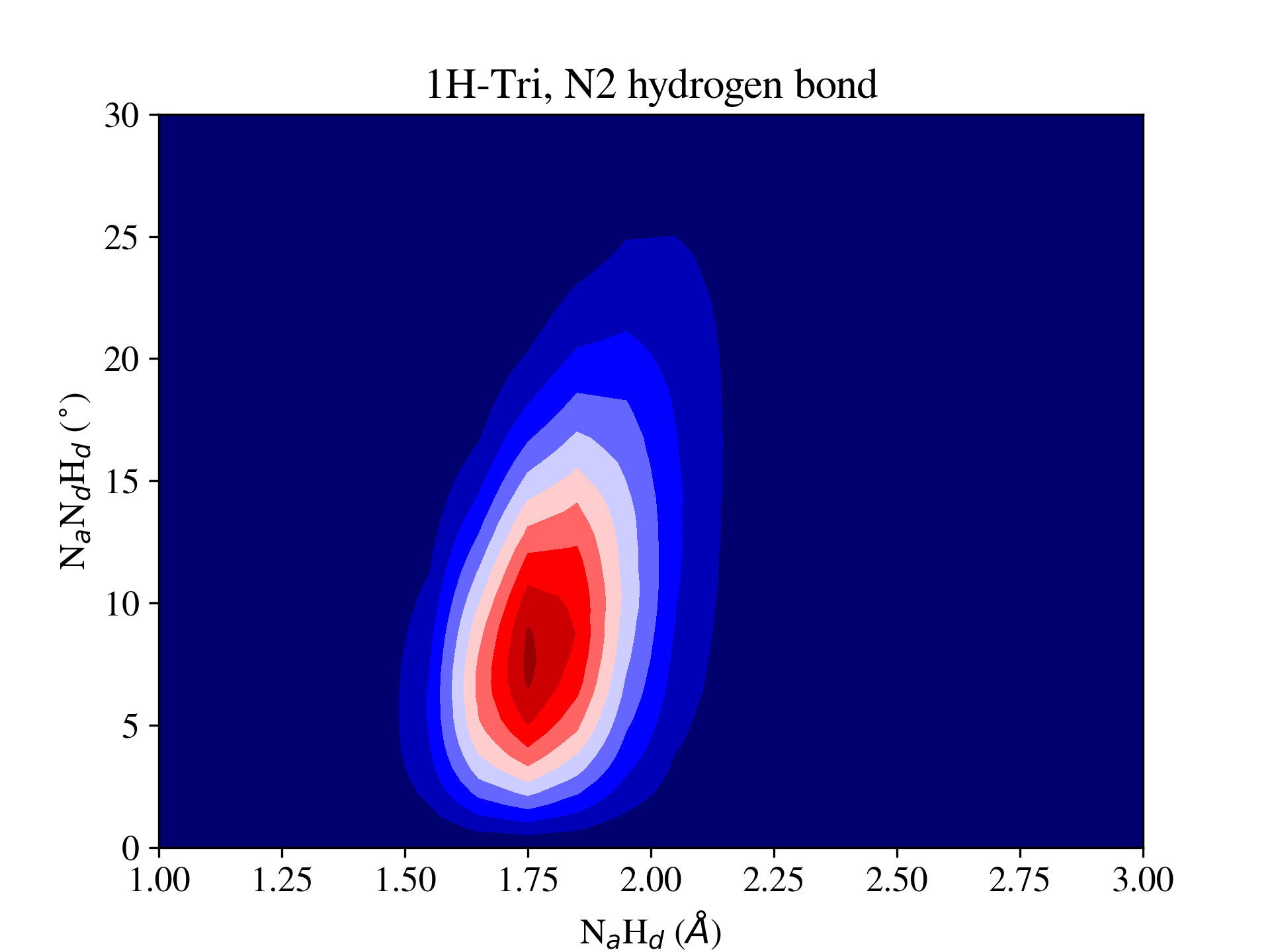}
   \end{minipage}
   \begin{minipage}{0.3\textwidth}
      \centering
      \includegraphics[width=\textwidth]{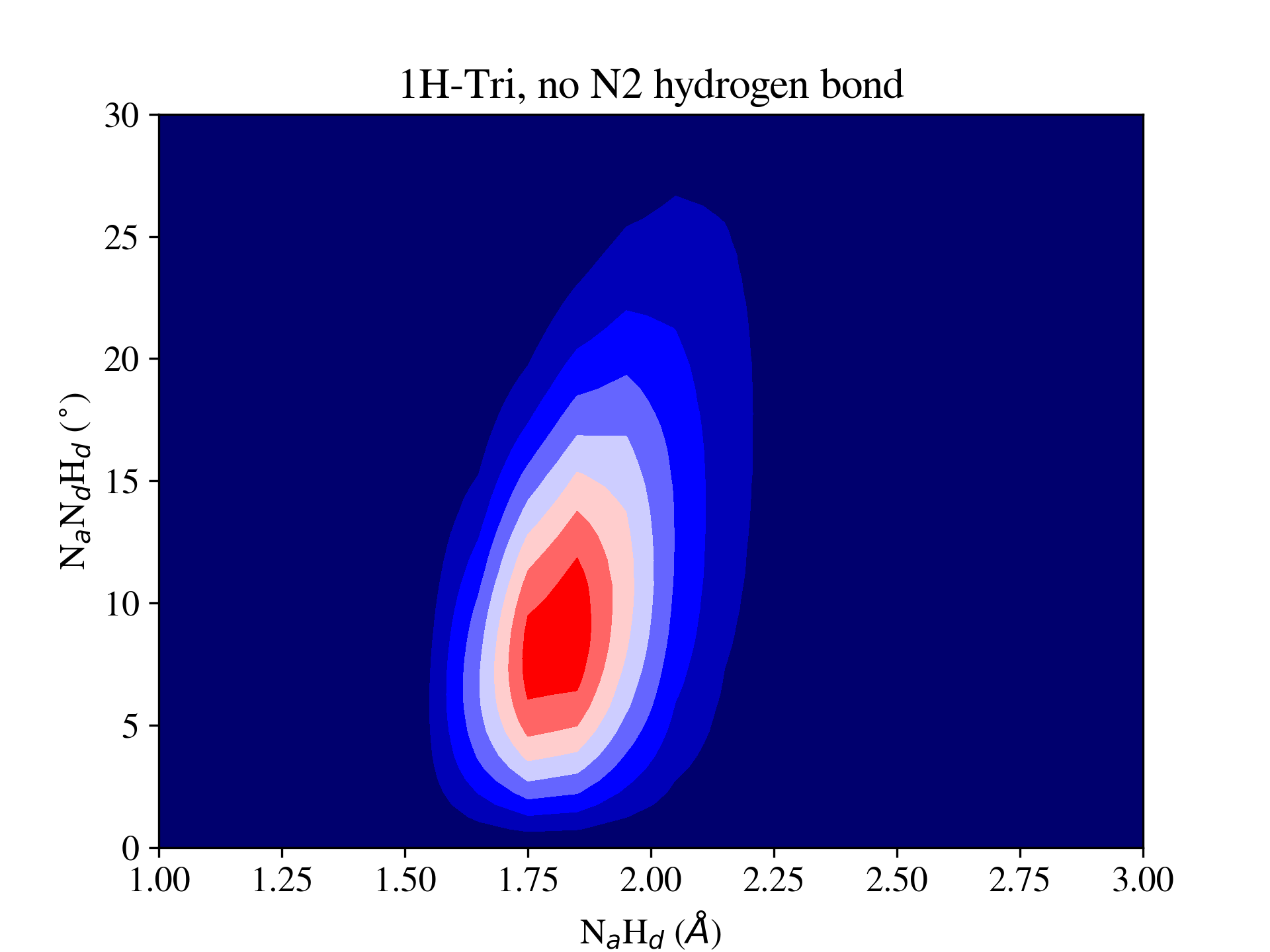}
   \end{minipage}
    \caption{Top panel: Hydrogen bond r-$\theta$ distributions for Imi, 1H-Tri, 2H-Tri, where r is the distance from the nearest hydrogen-bond-accepting nitrogen atom to the hydrogen-bond-donating hydrogen atom (N$_\textrm{a}$H$_\textrm{d}$), and $\theta$ is the angle formed between the N-H covalent bond and the vector connecting the hydrogen bond donor and acceptor nitrogen atoms(N$_\textrm{a}$N$_\textrm{d}$H$_\textrm{d}$). 
    Bottom panel: r-$\theta$ distributions for hydrogen bonds donated by N1 and N3 atoms in 1H-Tri, for molecules where the N2 atom accepts a hydrogen bond, and for molecules where it does not, respectively.}
    \label{fig:r_theta}
\end{figure*}

Finally, given the importance of the hydrogen bond at the N2 position in modulating the proton transfer rate in triazole, we consider its role in hydrogen bonding between molecules that do not hold the proton. Fig.~\ref{fig:r_theta} shows that when the N2 hydrogen bond is formed by a 1H-Tri molecule, it also acts to strengthen the hydrogen bonds formed at the N1 and N3 positions, indicated in the bottom panel by the tighter distribution of the hydrogen bond in distance and angle, leading to slower rearrangement. This leads to 1H-Tri possessing the slowest hydrogen bond relaxation time of 207~ps compared to 128~ps in 2H-Tri and 53~ps in Imi (see $\tau_3$ in SI Table~\ref{table:hbondcorr}). The ratio of these hydrogen bond relaxation times matches the ordering and the approximate ratios of their molecular diffusion constants ($D_{mol}$ in Table~\ref{table:msds}) where Imi diffuses $\sim6$ fold faster than 1H-Tri and $\sim3$ faster than 2H-Tri. This is consistent with hydrogen bond breaking being a precondition for diffusion of the molecules. For 1H-Tri, the 4-fold slower hydrogen-bond breaking than in Imi is close to its 6 fold difference in the proton diffusion coefficient, suggesting that some of the slowdown in the proton diffusion is accounted for by slower hydrogen bond rearrangements in molecules that surround the proton defect, with the rest arising from the ``blocking" mechanism. However, for 2H-Tri there is only a factor of 2 difference in the hydrogen bond lifetime compared with Imi but a 25 fold difference in their proton diffusion coefficients. This highlights that even for these chemically similar molecules, the hydrogen bond dynamics of the pure liquid are not always a reliable indicator of their ability to efficiently conduct protons. 

We have shown how the presence of the central nitrogen atom (N2) in 1H-Tri and 2H-Tri significantly alters their proton transport and hydrogen bond dynamics from that observed in Imi. In 2H-Tri, the covalent bond between N2 and hydrogen impedes proton transfer by limiting chain linearity and lowering the probability that a proton hop will result in 1,3-di-\textit{H}-1,2,3-triazolium, the preferred protonated tautomer. In 1H-Tri, hydrogen bonds received at the N2 position render a molecule unavailable to receive a proton, and thus limit the length of hydrogen bond chain available for the proton to traverse. Hydrogen bonds accepted at the N2 position also slow down the hydrogen bond chain reorganization in 1H-Tri by geometrically restricting the hydrogen bonds formed at the N1 and N3 positions, slowing down the rate at which protons can explore molecules not present in their original chains. These effects provide an explanation for the experimentally observed 10-fold faster proton diffusion in imidazole compared to 1,2,3-triazole.\cite{zhou2005promotion,kawada1970protonic} More generally, the insights obtained by the AIMD simulations reported here can be leveraged to design derivatized heterocycles for use as novel proton-conducting liquids exhibiting high proton transport rates.

\section*{acknowledgements}

This material is based upon work supported by the National Science Foundation Phase I CCI: NSF Center for First-Principles Design of Quantum Processes (CHE-1740645). T.E.M. also acknowledges support from the Camille Dreyfus Teacher-Scholar Awards Program. M.E.T. also acknowledges support from an Energy Frontiers Research Center funded by the U.S. Department of Energy, Office of Science, Basic Energy Sciences under Award no. DE-SC0019409

\clearpage
\onecolumngrid

\setcounter{equation}{0}
\setcounter{figure}{0}
\makeatletter 
\renewcommand{\thefigure}{S\@arabic\c@figure}
\renewcommand{\thetable}{S\@arabic\c@table}
\makeatother
\renewcommand{\theequation}{S\arabic{equation}}

\section*{Supporting Information}

\section{\label{sec:comp_details}Computational Details}

We performed classical \textit{ab initio} molecular dynamics simulations in the NVT ensemble with imidazole at 384 K and 1,2,3-triazole at 300K. Both simulation temperatures were a few degrees above the corresponding system's melting point. The simulations were conducted in cubic computational cells under periodic boundary conditions, with sides of length 19.337 $\r{A}$ for imidazole and 18.594 $\r{A}$ for 1,2,3-triazole. Every system contained 64 molecules, each with one excess proton. The simulation densities of imidazole (1.00 g/cm$^3$) and 1,2,3-triazole (1.14 g/cm$^3$) are consistent with their experimental values of 1.03 g/cm$^3$~\cite{lide2007crc} and 1.19 g/cm$^3$~\cite{pulst2016crystallization} respectively in the liquid phase. We performed 7 simulations of imidazole with lengths of 0.337 ns, 0.320 ns, 0.147 ns, 0.128 ns, 0.065 ns, 0.027 ns, and 0.024 ns, totaling 1.048 ns of trajectory. For imidazole, while dynamics were extracted from all trajectories, MSD values were calculated from the two longest trajectories and further analysis was based on the 5 longest ($\geq$ 0.065 ns in length). Three independent simulations were conducted for 1-$\textit{H}$-1,2,3-triazole (1H-Tri), with lengths of 0.457 ns, 0.523 ns, and 0.508 ns, totaling 1.488 ns, and three simulations were similarly conducted for 2-$\textit{H}$-1,2,3-triazole (2H-Tri), with lengths 0.551 ns, 0.494 ns, and 0.532 ns, totaling 1.577 ns. All the trajectories for both 1H-Tri and 2H-Tri were used to extract dynamics and perform subsequent analysis. Simulations were performed using the i-PI program~\cite{ceriotti2014pi} with an MTS integrator of the reversible reference system propagator algorithm (r-RESPA) type~\cite{tuckerman1992reversible} that utilized a 2.0 fs timestep for the full forces and a 0.5 fs timestep for the reference forces. Initial configurations were equilibrated for $\sim$4 ps using a local Langevin thermostat with a time constant of 25 fs. Production runs used a global stochastic velocity rescaling (SVR) thermostat~\cite{bussi2007canonical} with a time constant of 1 ps. The SVR thermostat couples to the total kinetic energy of the system, causing negligible disturbance to the dynamics of the trajectory.~\cite{ceriotti2010efficient} Full forces were evaluated using the CP2K program~\cite{hutter2014cp2k, vandevondele2005quickstep} at the revPBE level of DFT with D3 dispersion corrections~\cite{grimme2010consistent}. Core electrons were replaced by pseudopotentials of the Goedecker-Teter-Hutter type.~\cite{goedecker1996separable} Kohn-Sham orbitals were expanded in a TZV2P atom-centered basis set, while the density was expanded with a cutoff of 400 Ry. The MTS reference forces were evaluated at the self-consistent charge density-functional tight-binding (SCC-DFTB3) \cite{gaus2011dftb3} level of theory using the DFTB+ program~\cite{aradi2007dftb+}. The 3ob parameter was used for all atoms.~\cite{gaus2011dftb3} Dispersion forces were included through a Lennard-Jones potential~\cite{zhechkov2005efficient} with parameters taken from the Universal Force Field.~\cite{rappe1992uff}

\section{\label{sec:msd_plots}Mean Squared Displacement (MSD) Plots for the center of excess charge}
\begin{figure}[h!]
\centering
\includegraphics[width=0.5\textwidth]{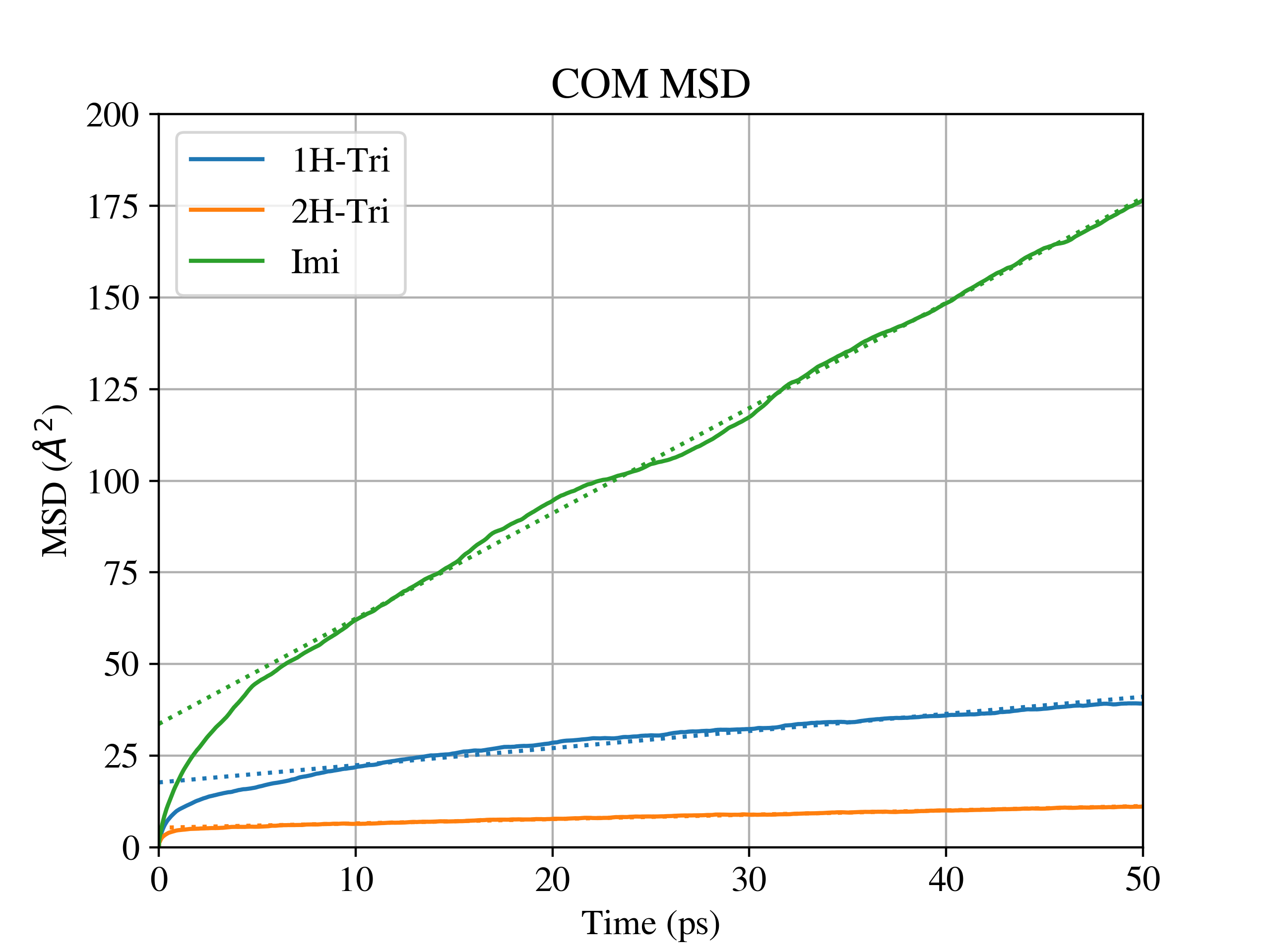}
\caption{Mean squared displacement plots for Imi, 1H-Tri, and 2H-Tri, with their accompanying linear fits shown as dotted lines.}
\label{fig:msds}
\end{figure}

The rate of proton diffusion was measured by tracking the center of excess charge (CEC), which is an imidazolium (1,2,3-triazolium) molecule that is determined by first assigning each acidic hydrogen atom (attached to nitrogen at frame 0) to its nearest nitrogen atom and then picking out the single imidazole (1,2,3-triazole) molecule with 2 acidic hydrogen atoms attached. Plots for the mean squared displacement (MSD) of the center of mass of the CEC are shown in Fig.~\ref{fig:msds} for imidazole(Imi), 1-$\textit{H}$-1,2,3-triazole(1H-Tri), and 2-$\textit{H}$-1,2,3-triazole(2H-Tri). Linear fits were conducted to obtain the proton diffusion coefficients reported in the main text, and error bars were set to one standard error of the mean over included trajectories.

\begin{figure}[h]
    \centering
    \begin{minipage}{0.3\textwidth}
       \centering
       \includegraphics[width=\textwidth]{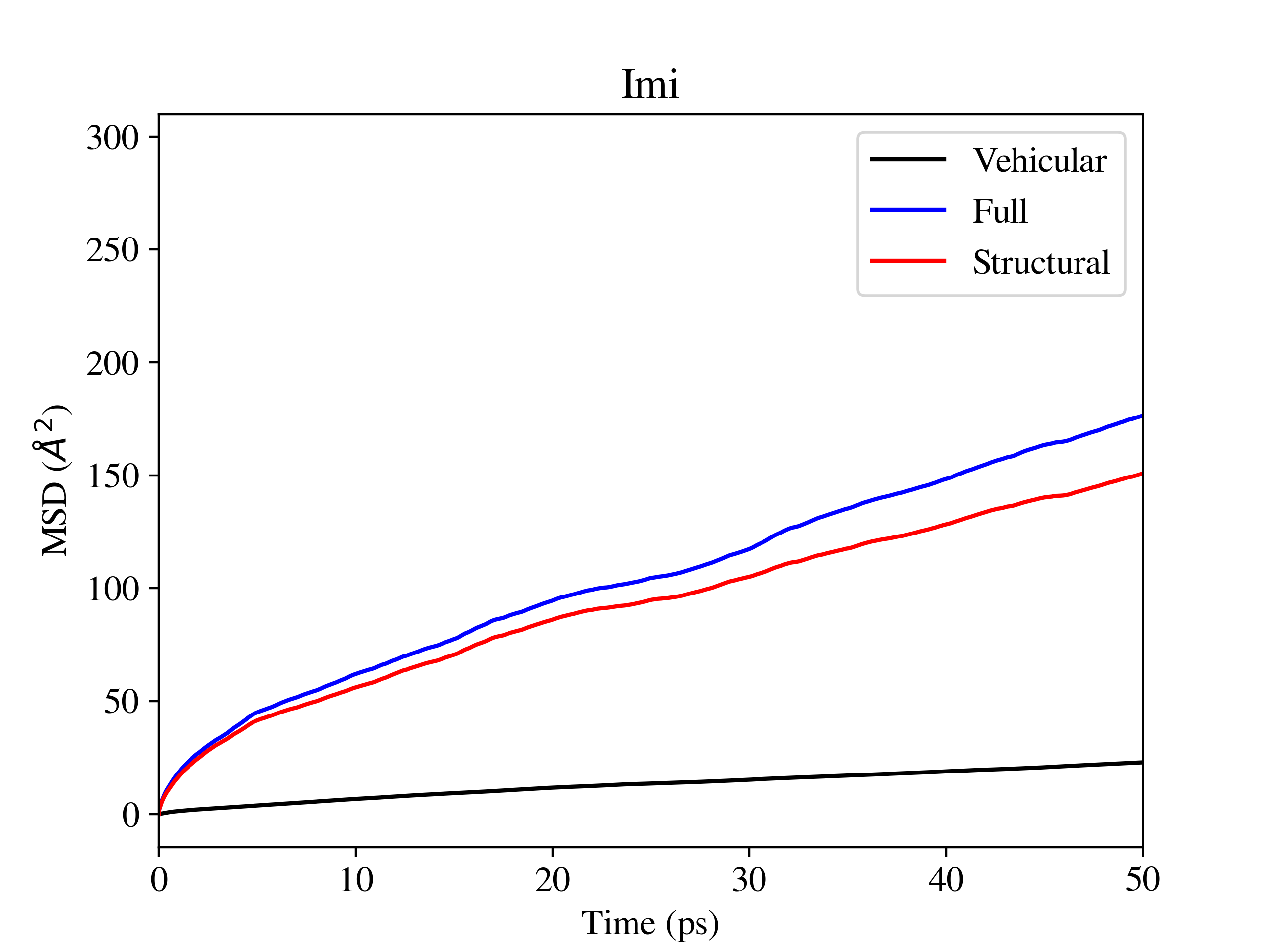}
   \end{minipage}
   \begin{minipage}{0.3\textwidth}
      \includegraphics[width=\textwidth]{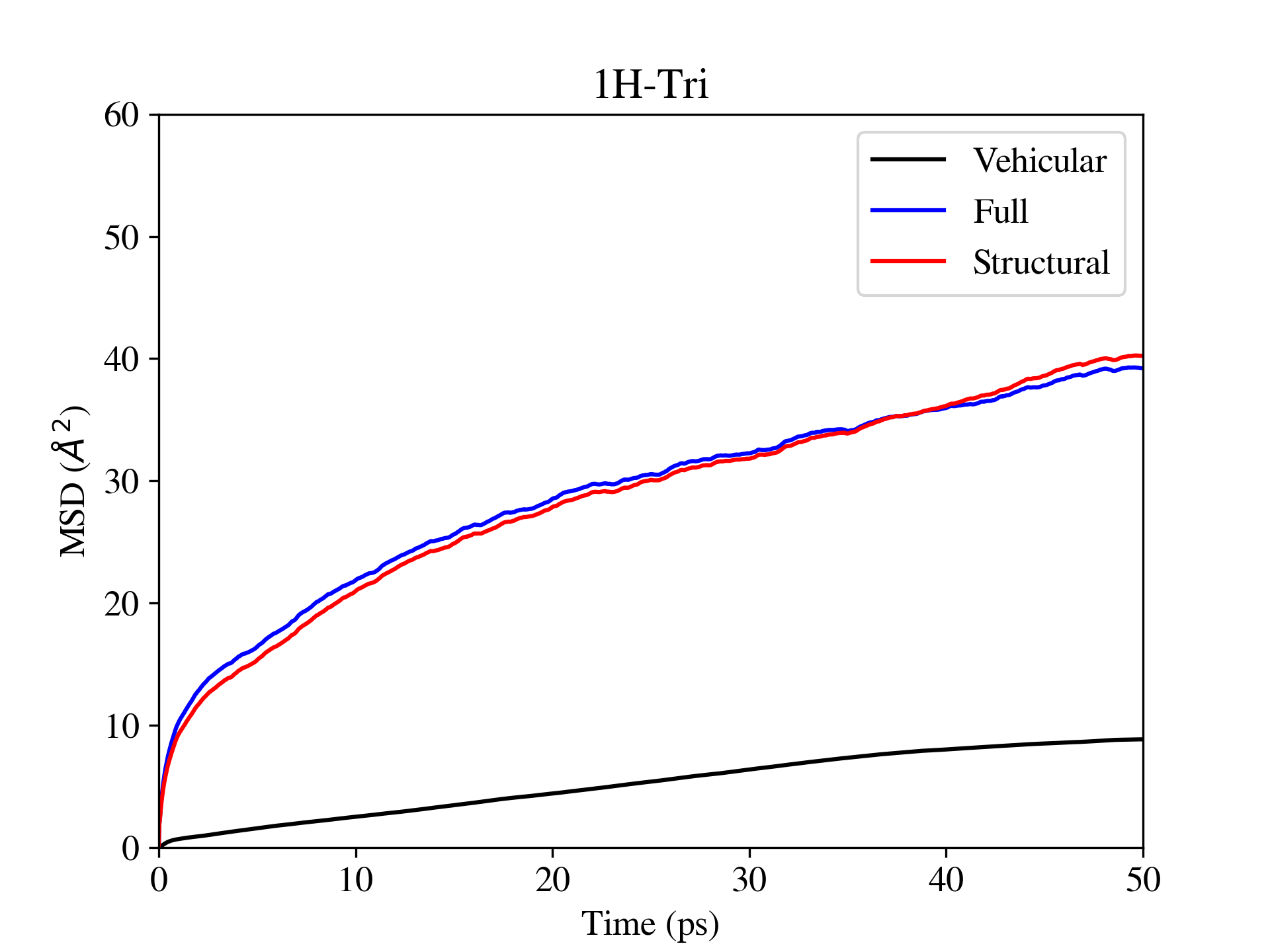}
   \end{minipage}
   \begin{minipage}{0.3\textwidth}
       \centering
       \includegraphics[width=\textwidth]{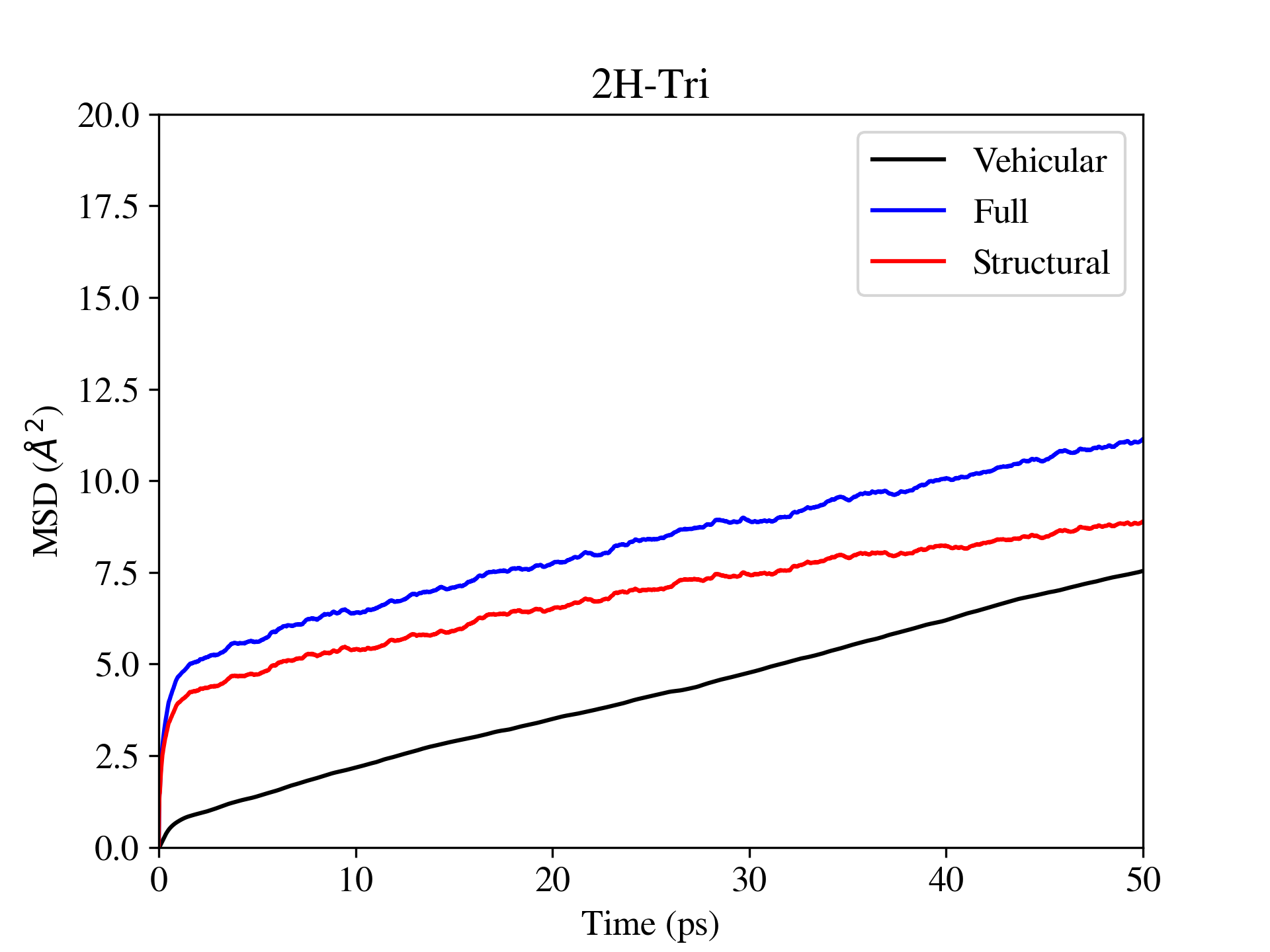}
   \end{minipage}
    \caption{Vehicular, structural, and full proton MSDs for Imi, 1H-Tri, and 2H-Tri. Vehicular and structural diffusion MSDs were obtained by splitting the full mean squared displacement of the proton into contributions from vehicular motion and intermolecular hops respectively.}
    \label{fig:decomposed-plots}
\end{figure}

The MSD of the proton was then decomposed by splitting proton displacements into those arising from intermolecular proton hops and those arising from vehicular motion, yielding the structural and vehicular MSDs respectively, which are shown in Figure~\ref{fig:decomposed-plots}. Vehicular and structural diffusion coefficients were then extracted from the MSDs, and their values are reported in Table~\ref{table:decomposition}.

\begin{table}[ht]
\setcellgapes{3pt}
\makegapedcells
\centering
\begin{tabularx}{\linewidth}{LRRR}
\toprule[1pt]
System & Vehicular diffusion coefficient ($\AA^2$/ps) & Structural diffusion Coefficient ($\AA^2$/ps) & Proton diffusion coefficient ($\AA^2$/ps)\\
\hline
Imidazole (384 K) & 0.077 $\pm$ 0.003 & 0.39 $\pm$ 0.04 & 0.47 $\pm$ 0.04\\
1-\textit{H}-1,2,3-triazole (300 K) & 0.022 $\pm$ 0.002 & 0.085 $\pm$ 0.002 & 0.08 $\pm$ 0.02\\
2-\textit{H}-1,2,3-triazole (300 K) & 0.032 $\pm$ 0.002 & 0.015 $\pm$ 0.007 & 0.020 $\pm$ 0.003 \\
\bottomrule[1pt]
\end{tabularx}
\caption{Vehicular, structural, and proton diffusion coefficients for Imi, 1H-Tri, and 2H-Tri. Vehicular and structural diffusion coefficients were obtained by splitting the mean squared displacement of the proton into contributions from molecular motion and intermolecular hops respectively.}
\label{table:decomposition}
\end{table}

\section{\label{hop_stats}Proton Transfer Statistics}
Table~\ref{table:hopstat} shows proton hop data obtained from our simulations, with an emphasis on productive proton hops, i.e. those that are not reversed by the next proton hop.

\begin{table}[ht]
\setcellgapes{3pt}
\makegapedcells
\centering
\begin{tabularx}{\linewidth}{LRRRRR}
\toprule[1pt]
System & Trajectory & Total number of hops & Number of productive proton hops (no rattling) & Fraction of productive proton hops (no rattling) & Length of trajectory (ps)\\
\hline
\multirow{5}{*}{Imidazole (384 K)} & 1 & 1351 & 115 & 0.085 & 320 \\
                                   & 2 &  305 &  24 & 0.089 &  65 \\
                                   & 3 & 1455 & 111 & 0.076 & 337 \\
                                   & 4 &  500 &  40 & 0.080 & 147 \\
                                   & 5 &  666 &  50 & 0.075 & 128 \\
\hline
\multirow{3}{*}{1-\textit{H}-1,2,3-triazole (300 K)} & 1 & 1892 & 44 & 0.023 & 457\\
                                                     & 2 & 2627 & 87 & 0.033 & 523\\
                                                     & 3 & 1582 & 34 & 0.021 & 508\\
\hline
\multirow{3}{*}{2-\textit{H}-1,2,3-triazole (300 K)} & 1 & 3451 & 47 & 0.014 & 551\\
                                                     & 2 & 1563 &  7 & 0.004 & 494\\
                                                     & 3 &  199 &  5 & 0.025 & 532\\
\hline
\end{tabularx}
\caption{Proton hop statistics for Imi, 1H-Tri, and 2H-Tri.}
\label{table:hopstat}
\end{table}

\section{\label{sec:bond_overview}Hydrogen bond data for Imidazole and 1,2,3-triazole}
\begin{figure}[h!]
    \centering
    \includegraphics[width=0.5\textwidth]{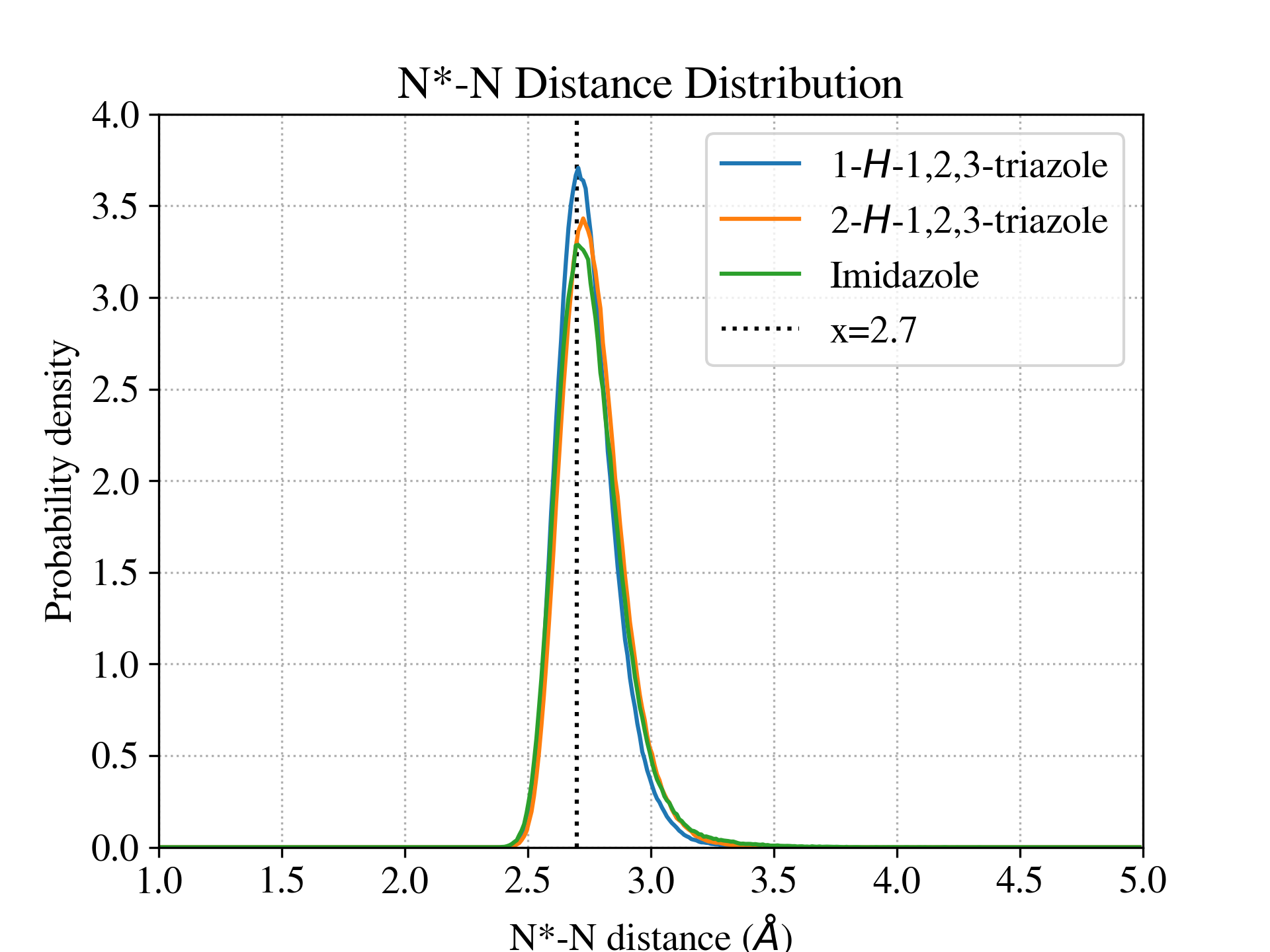}
    \caption{Probability distributions for the N*-N distances for Imi, 1H-Tri, and 2H-Tri}
    \label{fig:NsN_dists}
\end{figure}

We compared hydrogen bonds donated by protonated molecules of imidazole and 1,2,3-triazole by first plotting the N* - N length distribution, where N* refers to the two protonated nitrogen atoms on the center of excess charge and N refers to the adjacent N atoms in the first solvation shell. This distribution is shown in Fig.~\ref{fig:NsN_dists}. The peak N* - N length for 1H-Tri was used to calculate the steady-state intra-chain mean squared displacement (MSD($\inf$)) as described in the main text.

\begin{figure}[h]
    \centering
    \includegraphics[width=0.5\textwidth]{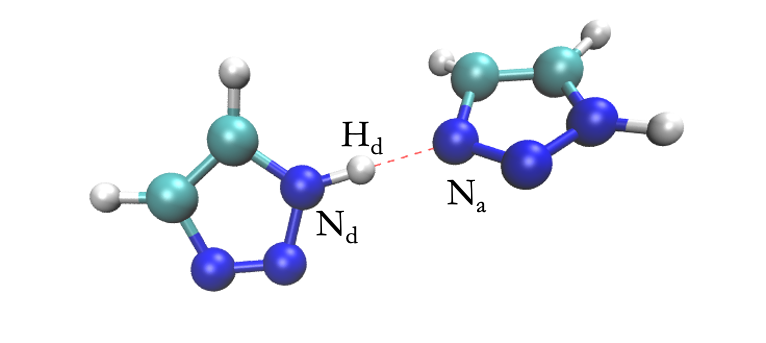}
    \caption{Atomic labels for hydrogen bond participants, illustrated here using two adjacent molecules of }
    \label{fig:h-bond-illustrated}
\end{figure}

\begin{figure}[h]
    \centering
    \begin{minipage}{0.3\textwidth}
       \centering
       \includegraphics[width=\textwidth]{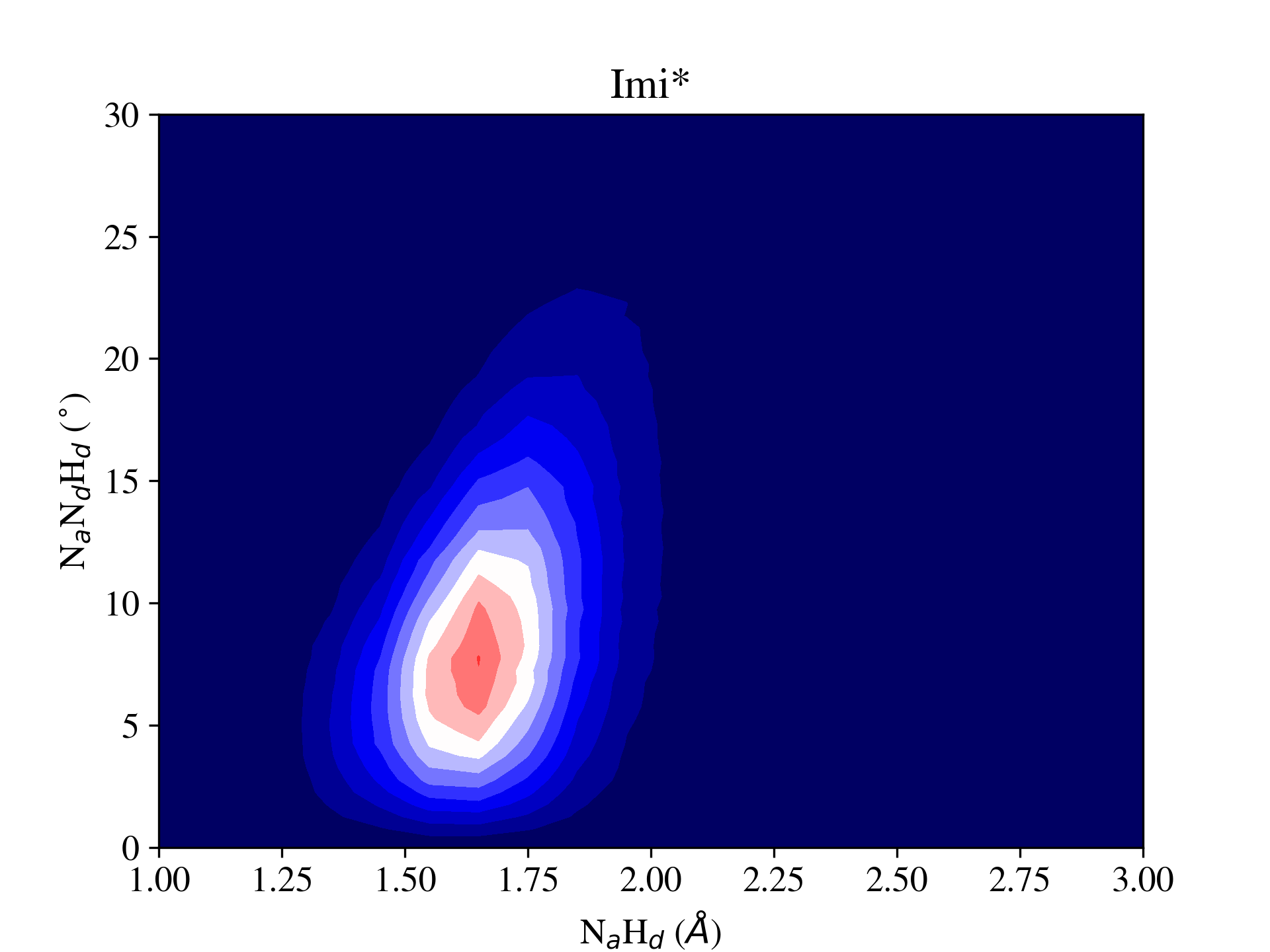}
   \end{minipage}
   \begin{minipage}{0.3\textwidth}
      \includegraphics[width=\textwidth]{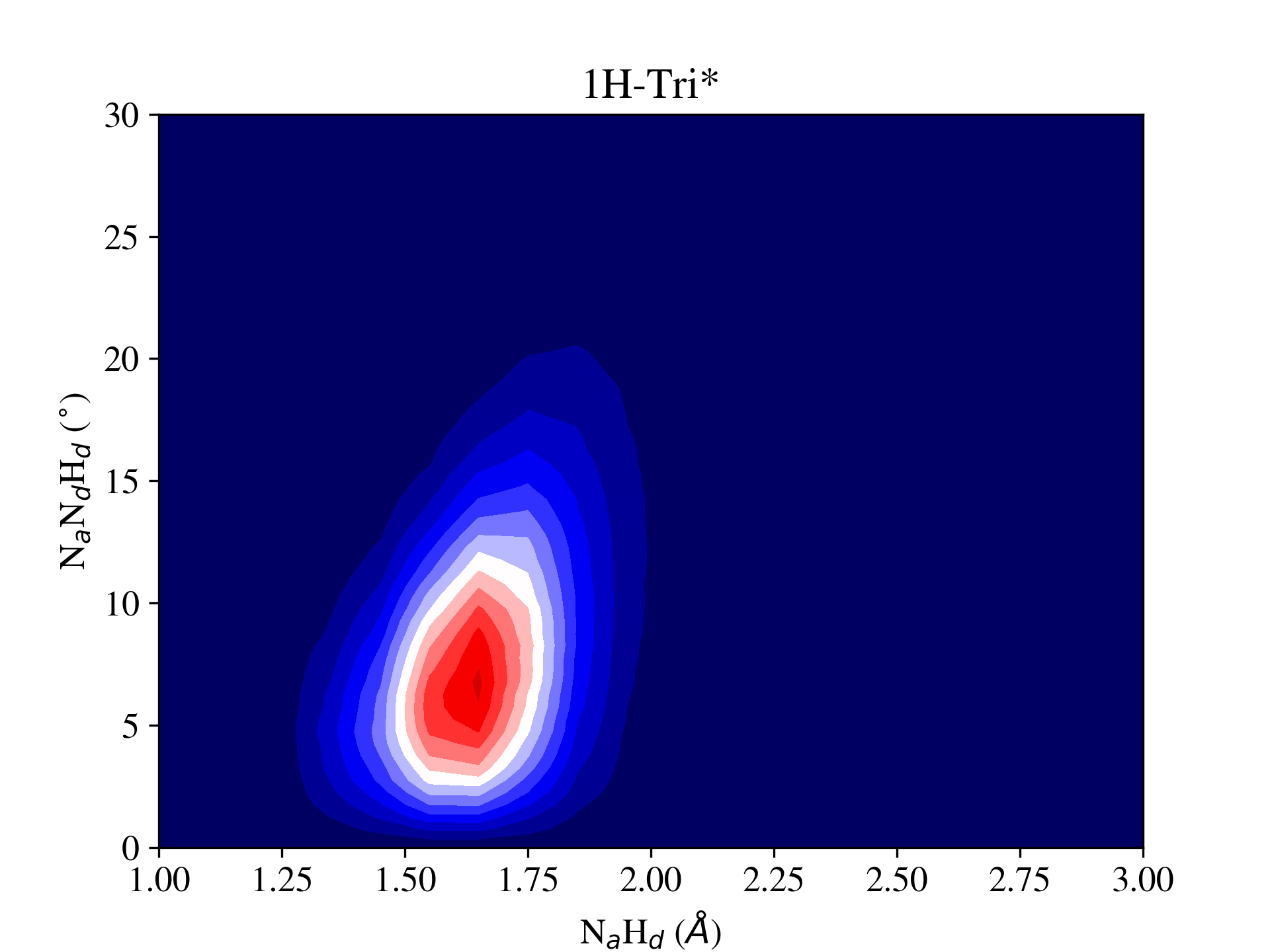}
   \end{minipage}
   \begin{minipage}{0.3\textwidth}
       \centering
       \includegraphics[width=\textwidth]{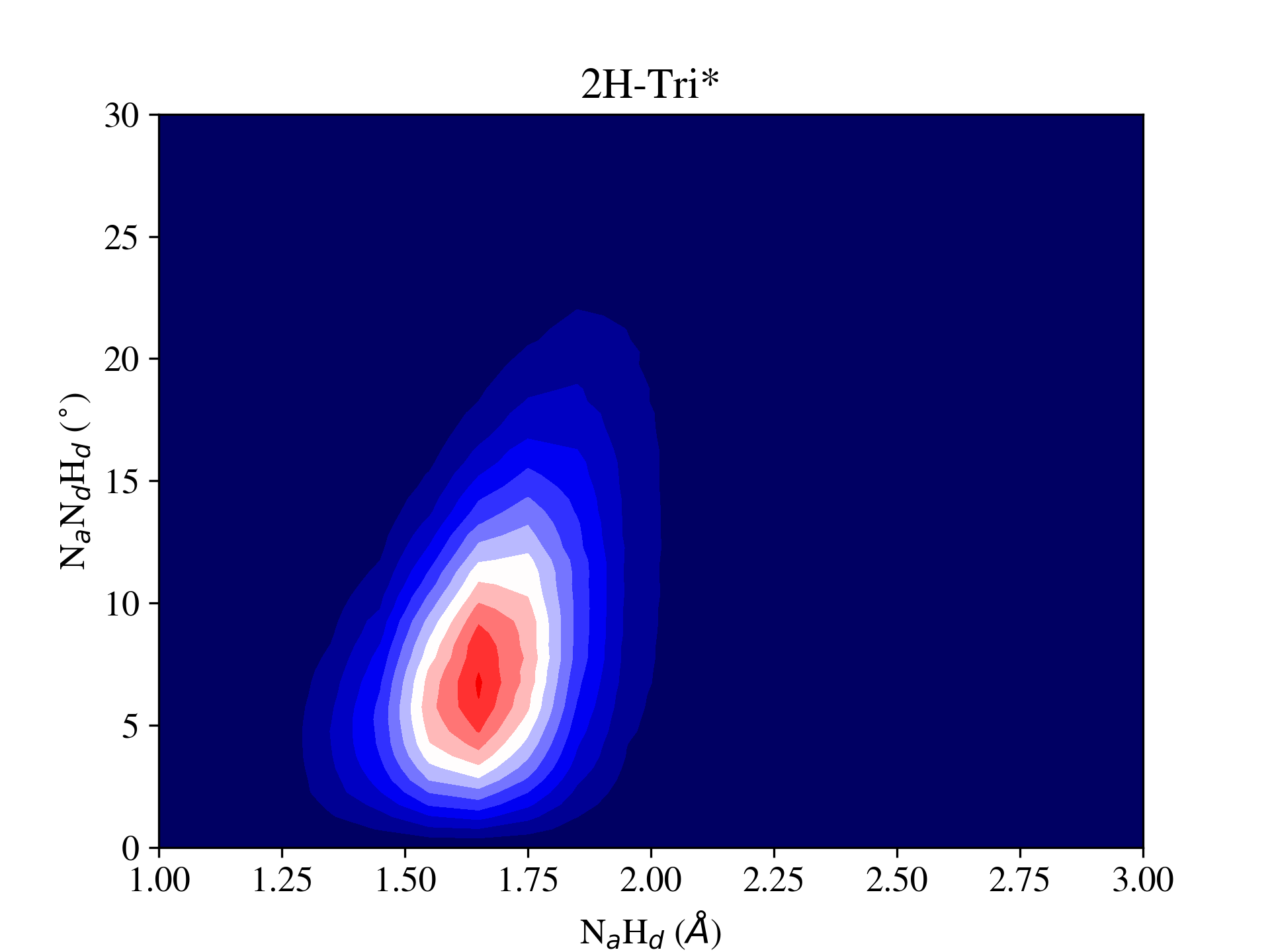}
   \end{minipage}
   \begin{minipage}{0.3\textwidth}
       \centering
       \includegraphics[width=\textwidth]{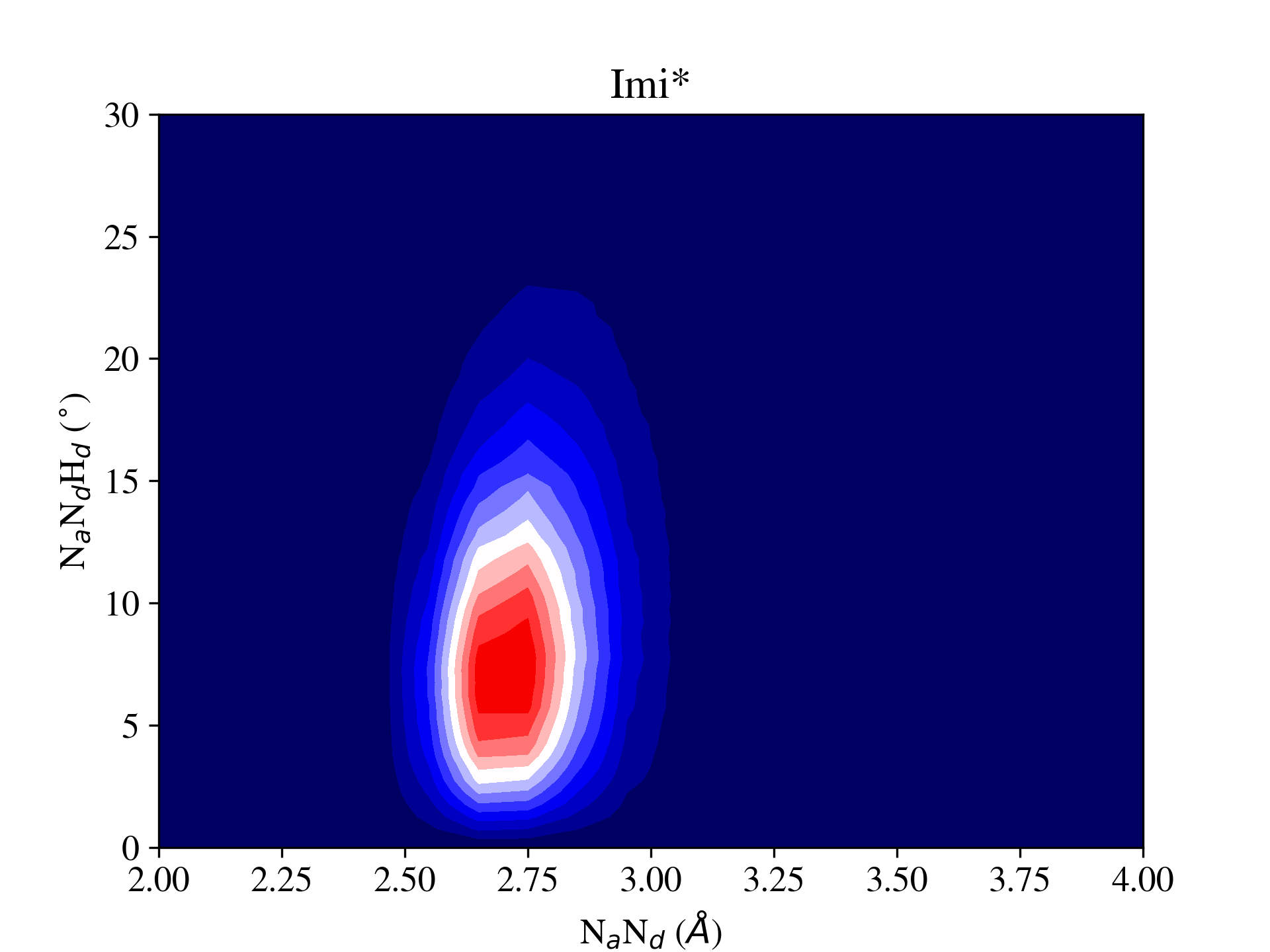}
   \end{minipage}
   \begin{minipage}{0.3\textwidth}
      \includegraphics[width=\textwidth]{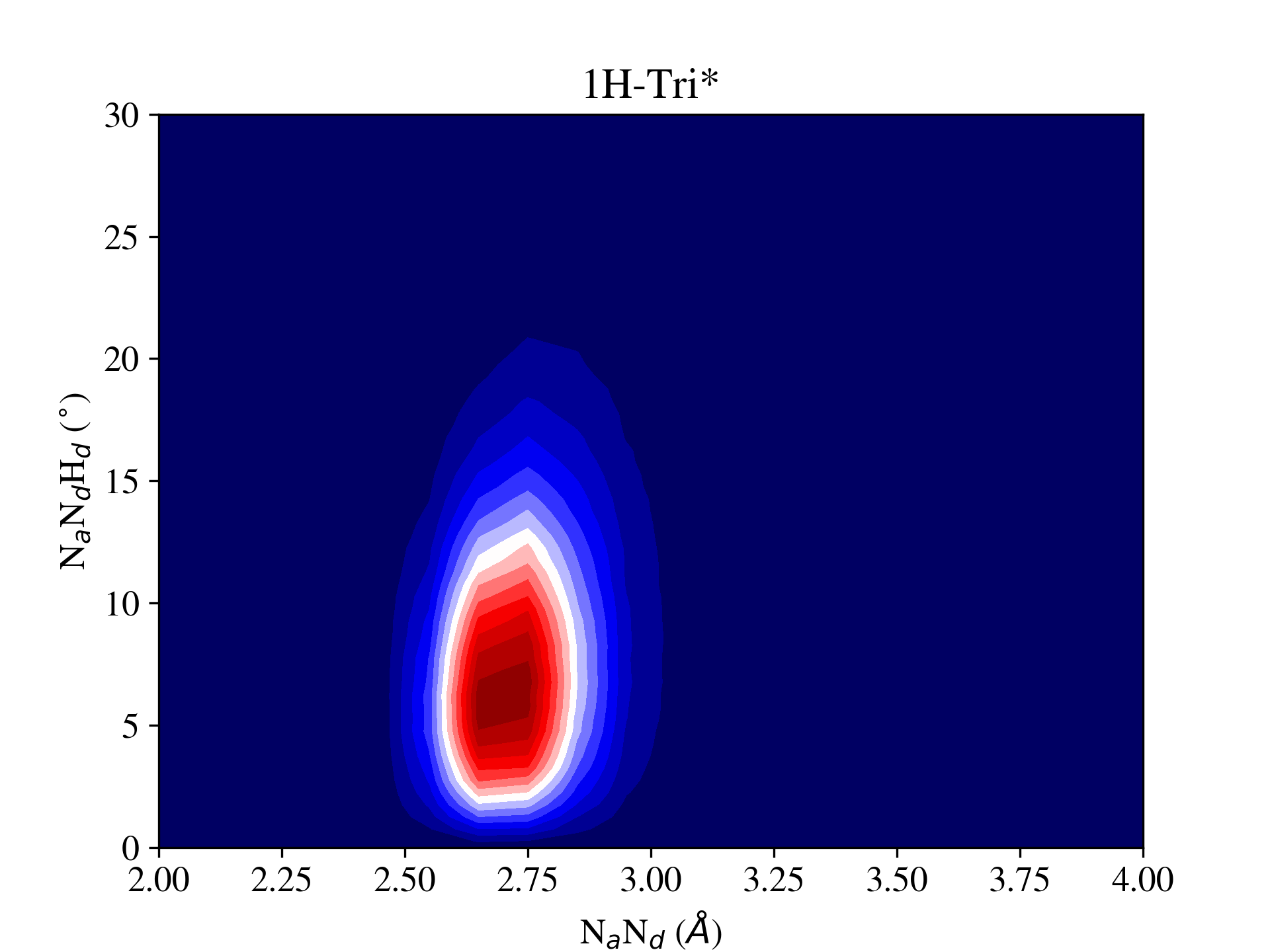}
   \end{minipage}
   \begin{minipage}{0.3\textwidth}
       \centering
       \includegraphics[width=\textwidth]{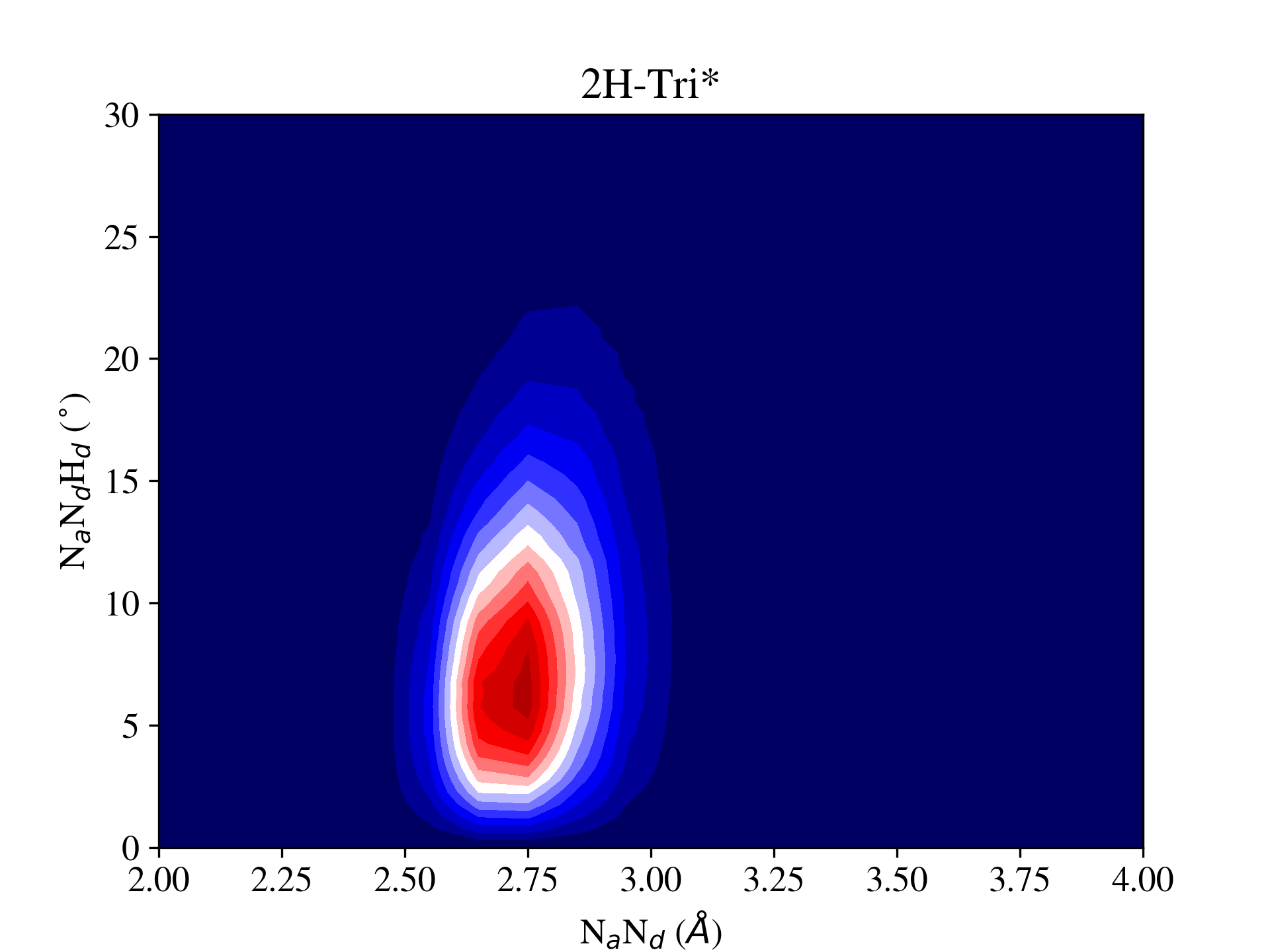}
   \end{minipage}
    \caption{Top panel: Hydrogen bond r-$\theta$ distributions for Imi*, 1H-Tri*, 2H-Tri*, where r = N$_\textrm{a}$N$_\textrm{d}$ and $\theta$ = N$_\textrm{a}$H$_\textrm{d}$H$_\textrm{d}$ as illustrated in Fig.~\ref{fig:h-bond-illustrated}.\\ Bottom panel: Hydrogen bond r-$\theta$ distributions for Imi*, 1H-Tri*, 2H-Tri*, where r = N$_\textrm{a}$N$_\textrm{d}$ and $\theta$ = N$_\textrm{a}$N$_\textrm{d}$H$_\textrm{d}$ as illustrated in Fig.~\ref{fig:h-bond-illustrated}.}
    \label{fig:protonated data}
\end{figure}

A more comprehensive picture of the hydrogen bonds donated by protonated molecules in imidazole and 1,2,3-triazole is shown in Fig.~\ref{fig:protonated data}, which contains distance-angle probability distributions for protonated molecules and their closest intermolecular nitrogen atoms. The atomic labels used to characterize the distances and angles of hydrogen bonds are shown in Fig.~\ref{fig:h-bond-illustrated}, where N$_\textrm{d}$ and H$_\textrm{d}$ refer to the hydrogen-bond-donating nitrogen atom and its covalently bonded hydrogen atom respectively, while N$_\textrm{a}$ refers to the hydrogen-bond-accepting nitrogen atom. In the distance-angle probability plots, $r$ refers to the lengths N$_\textrm{a}$H$_\textrm{d}$ and N$_\textrm{a}$N$_\textrm{d}$ in the top and bottom panels respectively, while $\theta$ refers to the angle N$_\textrm{a}$N$_\textrm{d}$H$_\textrm{d}$.

\begin{figure*}
    \centering
    \begin{minipage}{0.3\textwidth}
       \centering
       \includegraphics[width=\textwidth]{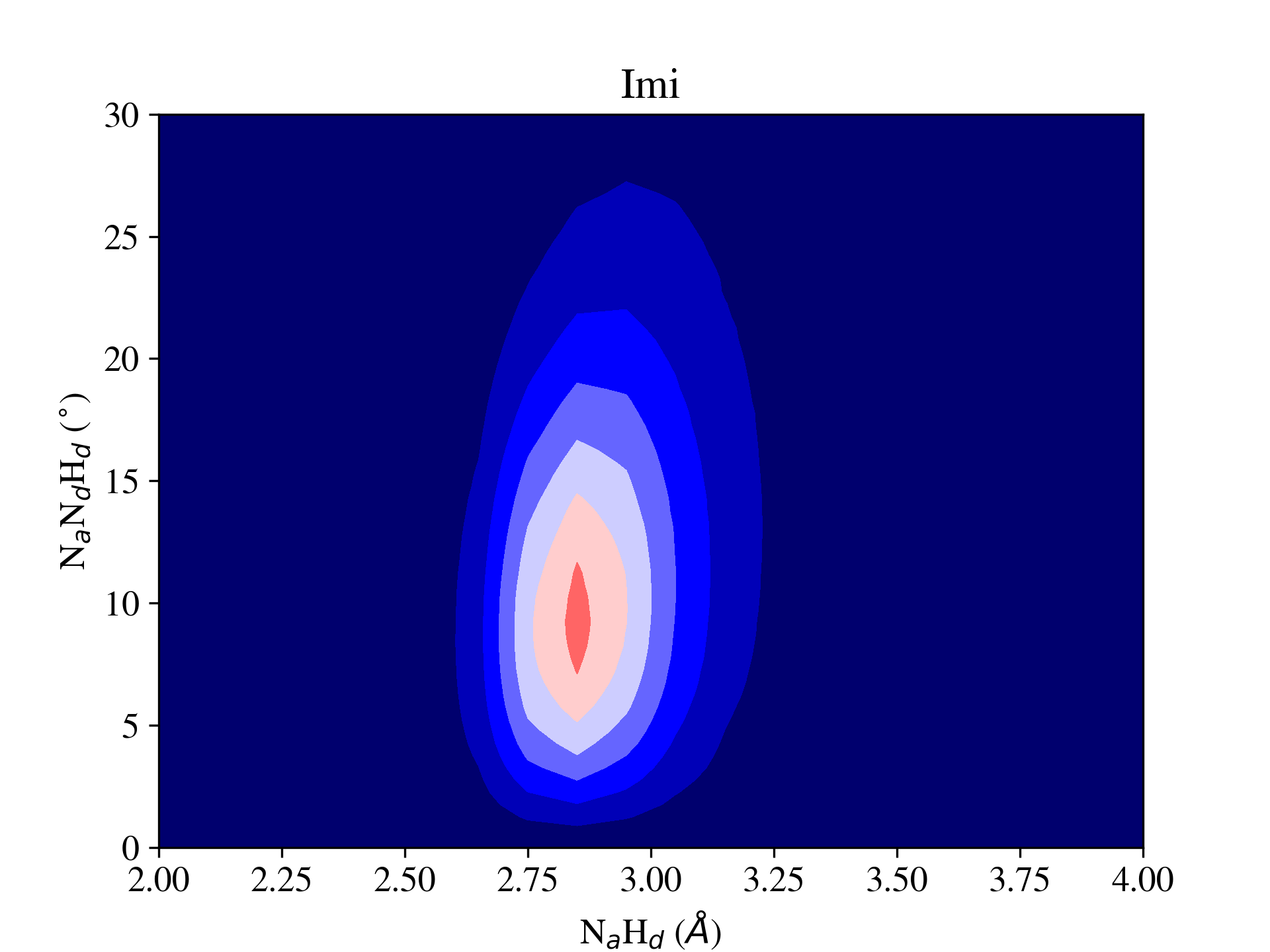}
   \end{minipage}
   \begin{minipage}{0.3\textwidth}
      \includegraphics[width=\textwidth]{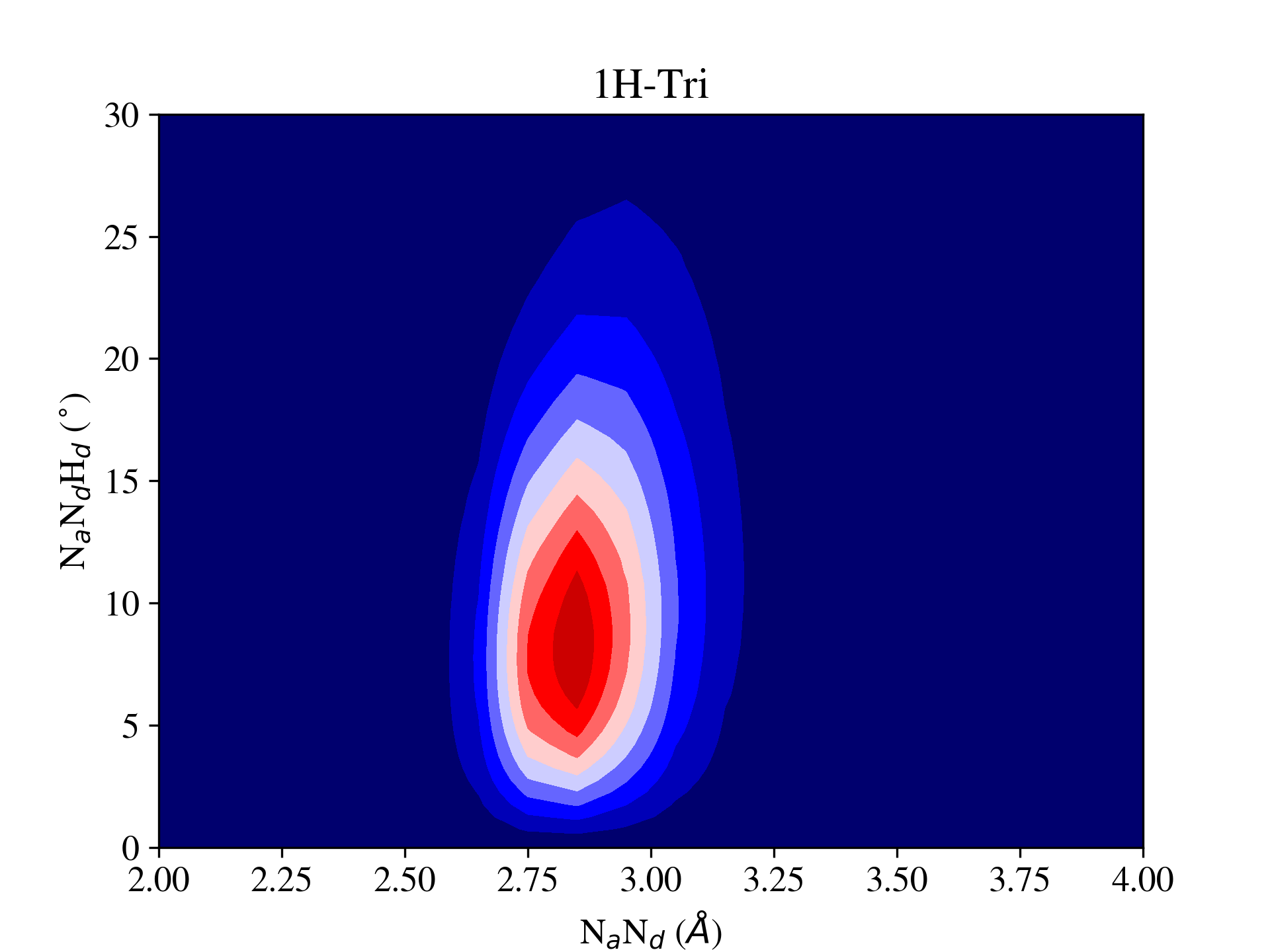}
   \end{minipage}
   \begin{minipage}{0.3\textwidth}
       \centering
       \includegraphics[width=\textwidth]{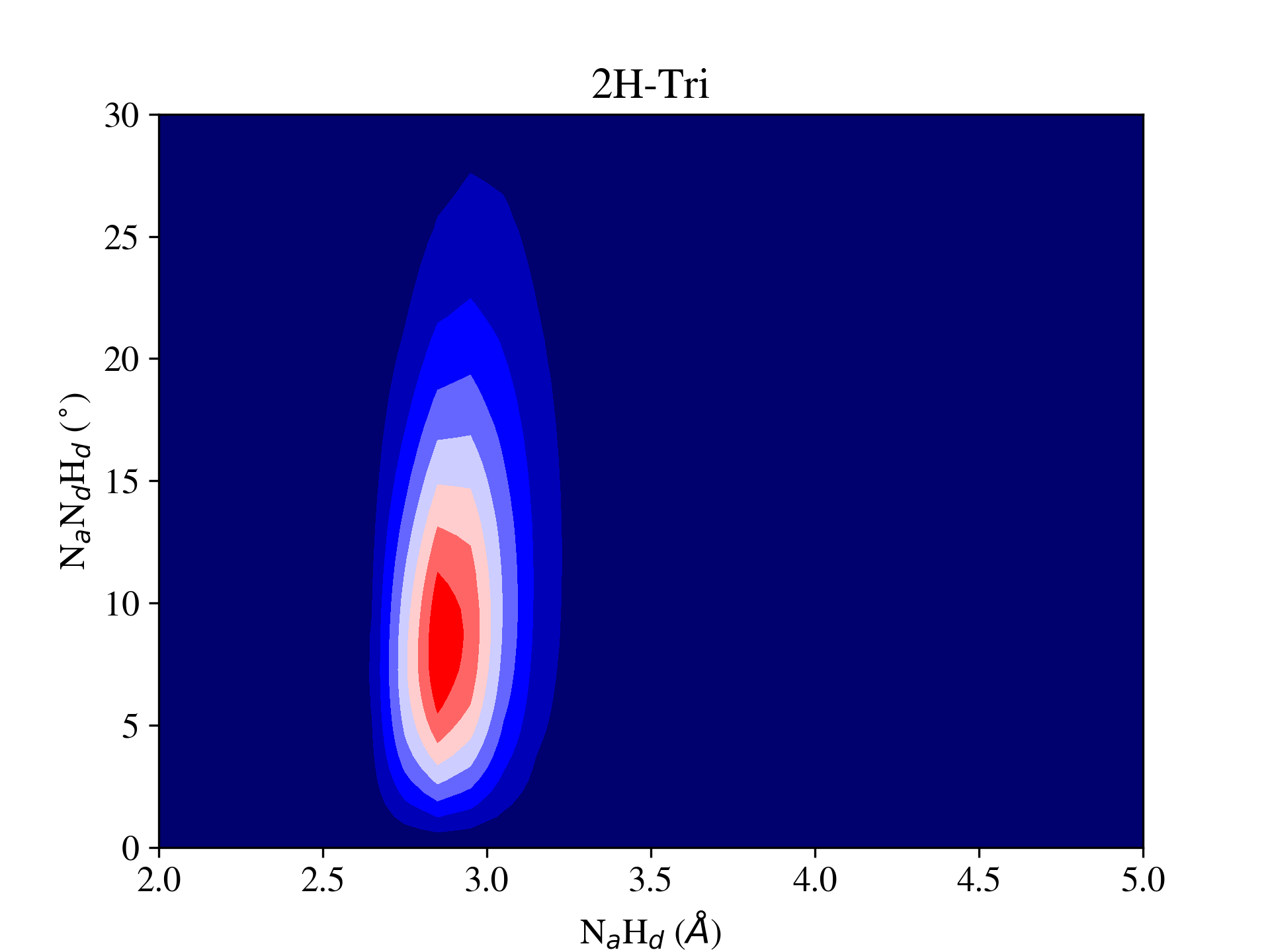}
   \end{minipage}
   \begin{minipage}{0.3\textwidth}
       \centering
       \includegraphics[width=\textwidth]{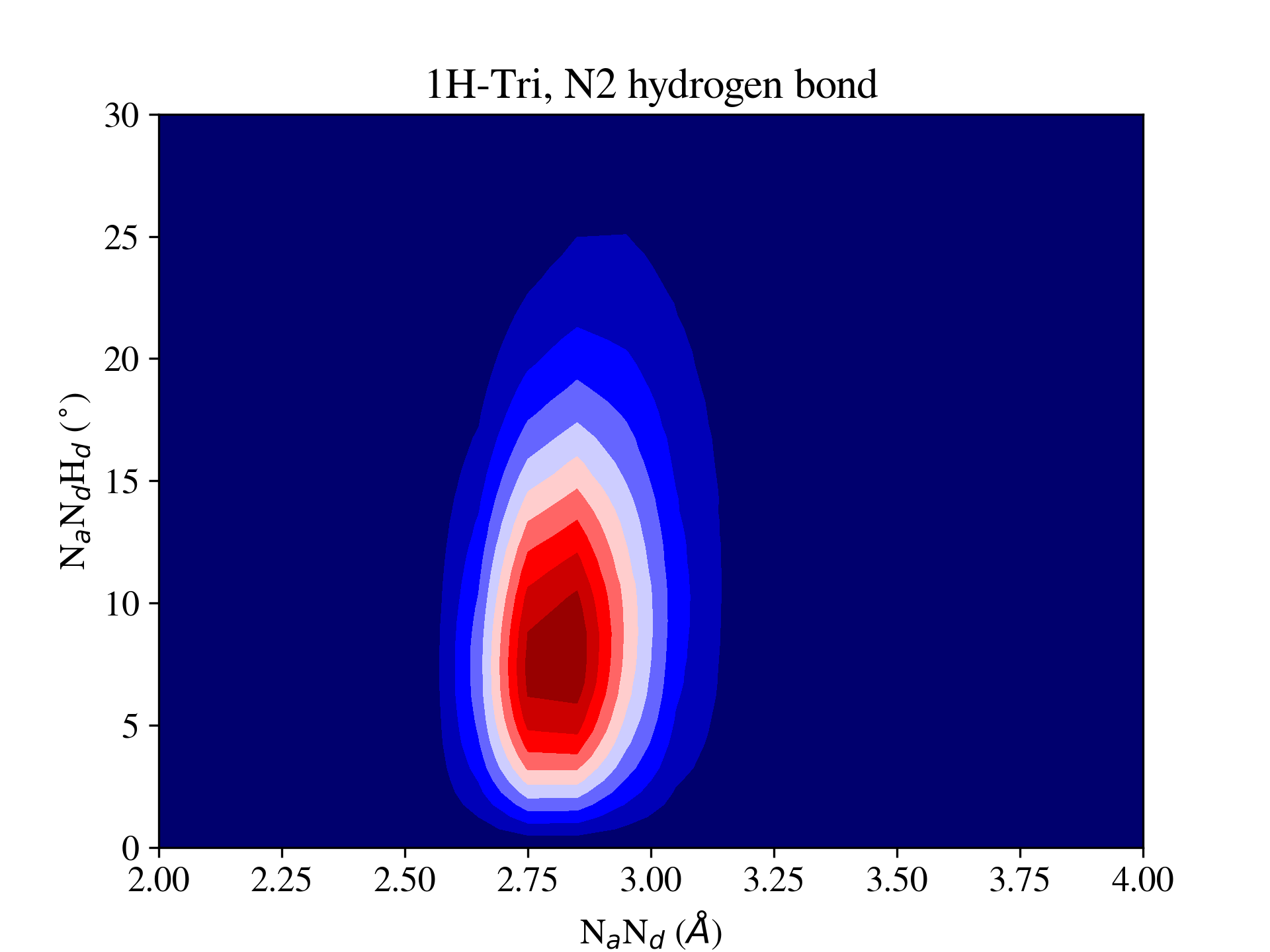}
   \end{minipage}
   \begin{minipage}{0.3\textwidth}
      \centering
      \includegraphics[width=\textwidth]{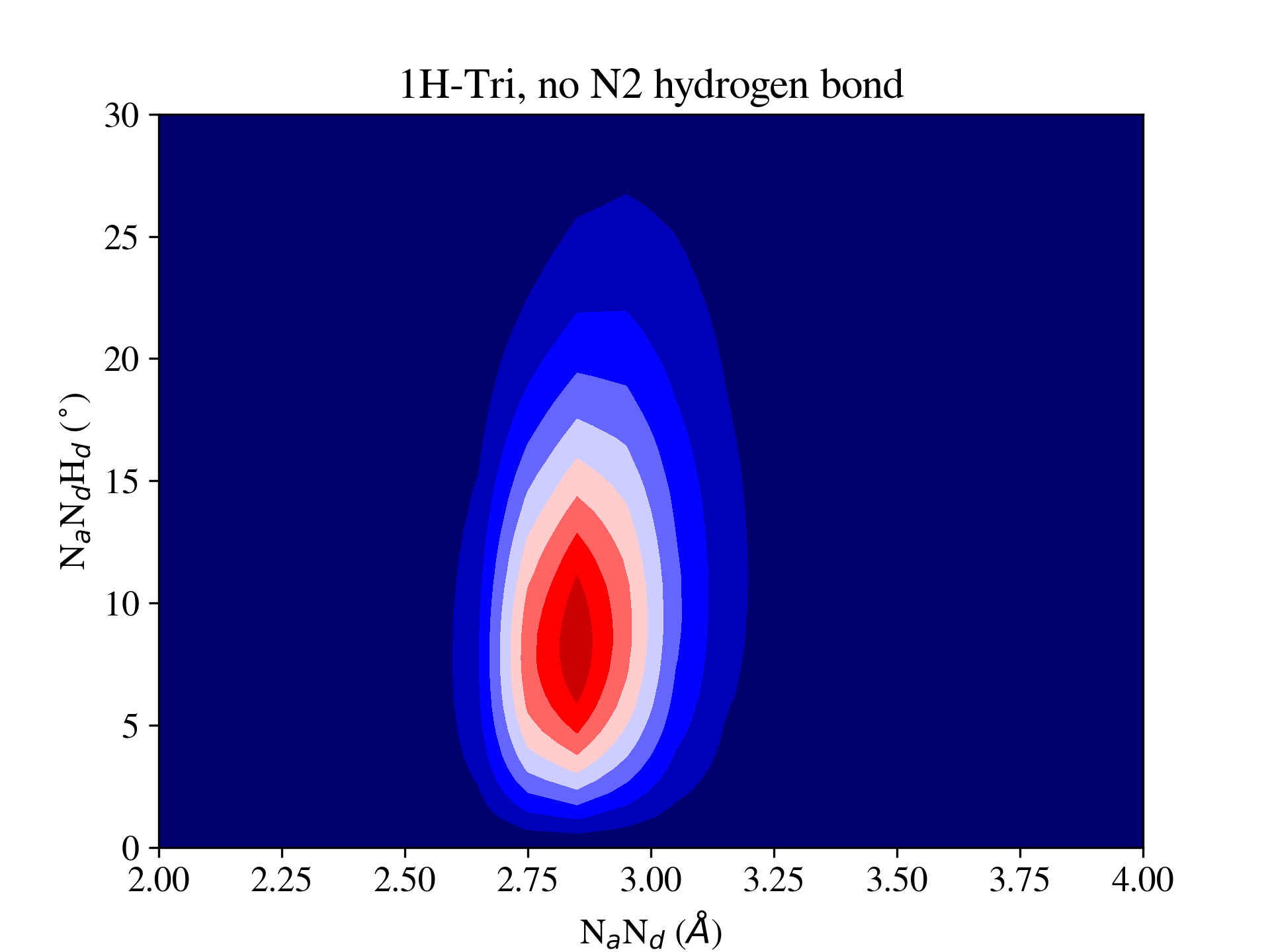}
   \end{minipage}
    \caption{Top panel: Hydrogen bond r-$\theta$ distributions for Imi, 1H-Tri, 2H-Tri, where r = N$_\textrm{a}$N$_\textrm{d}$ and $\theta$ = N$_\textrm{a}$N$_\textrm{d}$H$_\textrm{d}$ as illustrated in Fig.~\ref{fig:h-bond-illustrated}. 
    Bottom panel: r-$\theta$ distributions of hydrogen bonds donated by N1 and N3 atoms in 1H-Tri, for molecules where the N2 atom accepts a hydrogen bond, and for molecules where it does not, respectively.}
    \label{fig:r_theta_N}
\end{figure*}

Fig.~\ref{fig:r_theta_N} shows the distance-angle probability distributions for intermolecular nitrogen atoms nearest to hydrogen-bearing nitrogen atoms in imidazole and 1,2,3-triazole for molecules that are {\it not protonated}. Here, $r$=N$_\textrm{a}$N$_\textrm{d}$.

\newpage
\section{Protonation population correlation functions}

The protonation population function formalism introduced by Chandra et al \cite{Chandra2007,Tuckerman2010} can be used to study the timescales of proton transfer. We computed "intermittent" protonation population functions, which give the probability that a molecule will be protonated at time t if it was protonated at t=0. The population functions are then fit to a triexponential of the form:

\begin{equation}
    \sum_{i=1}^3a_ie^{-t/\tau_i}
\end{equation}

\begin{figure}[h]
    \centering
    \includegraphics[width=0.5\textwidth]{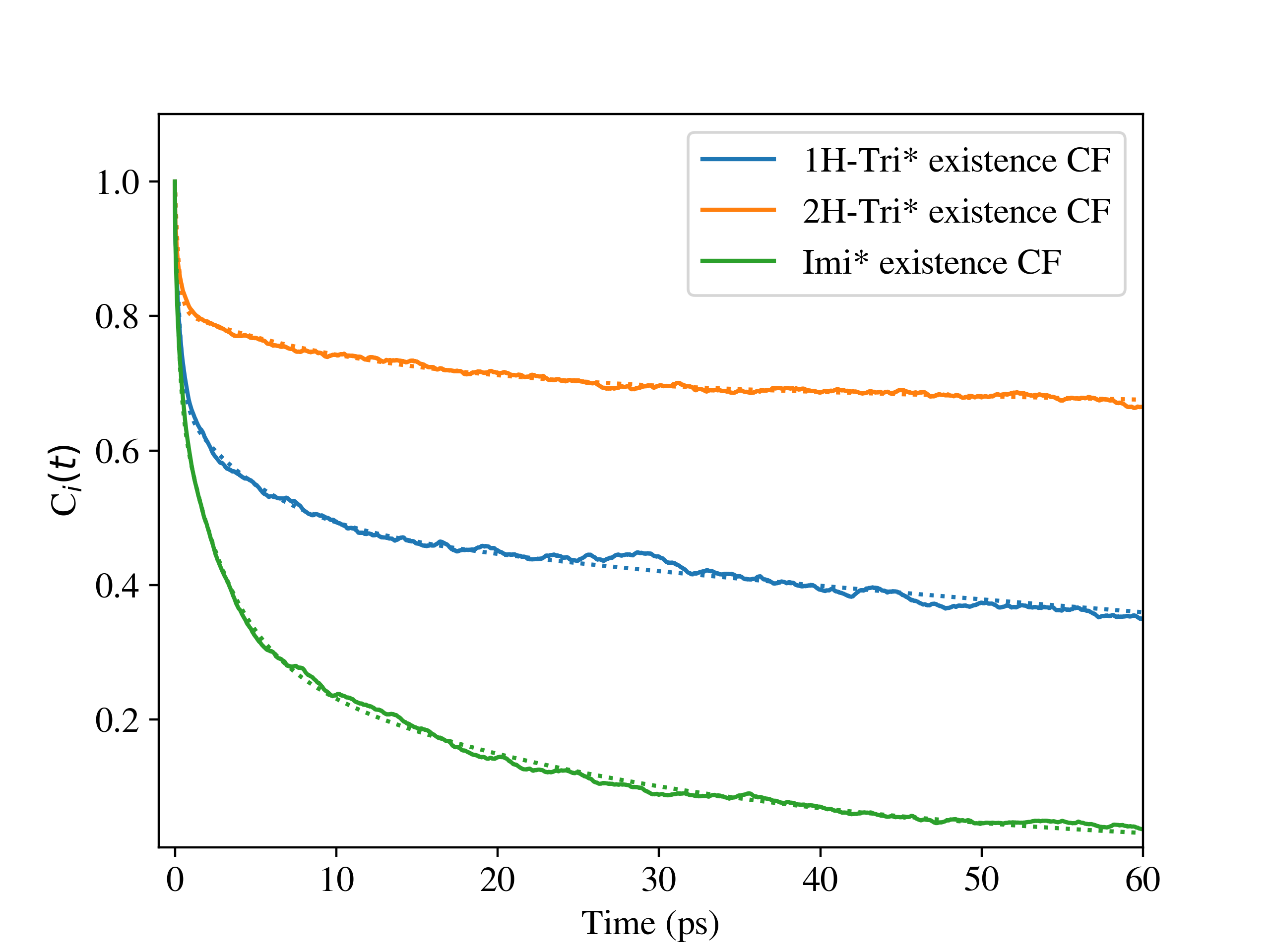}
    \caption{Protonation population correlation functions for Imi, 1H-Tri, and 2H-Tri. Their corresponding triexponential fits are shown as dotted lines.}
    \label{fig:corr_funs}
\end{figure}

The population functions and their resulting triexponential fits are shown in Fig.~\ref{fig:corr_funs} and Table~\ref{table:popcorr}.

\begin{table}[ht]
\setcellgapes{3pt}
\makegapedcells
\centering
\begin{tabularx}{\linewidth}{LRRR}
\toprule[1pt]
System & $\tau_1$ & $\tau_2$ & $\tau_3$\\
\hline
Imidazole (384 K) & 0.213 & 2.85 & 25.6 \\
1-\textit{H}-1,2,3-triazole (300 K) & 0.300 & 5.22 & 193 \\
2-\textit{H}-1,2,3-triazole (300 K) & 0.245 & 12.0 & 1604* \\
\bottomrule[1pt]
\end{tabularx}
\caption{Time constants for the protonation population function in Imi, 1H-Tri, and 2H-Tri\\ * - not converged; timescale exceeds length of trajectory.}
\label{table:popcorr}
\end{table}

\newpage
\section{Hydrogen bond population correlation functions}
\begin{figure}[h]
    \centering
    \includegraphics[width=0.5\textwidth]{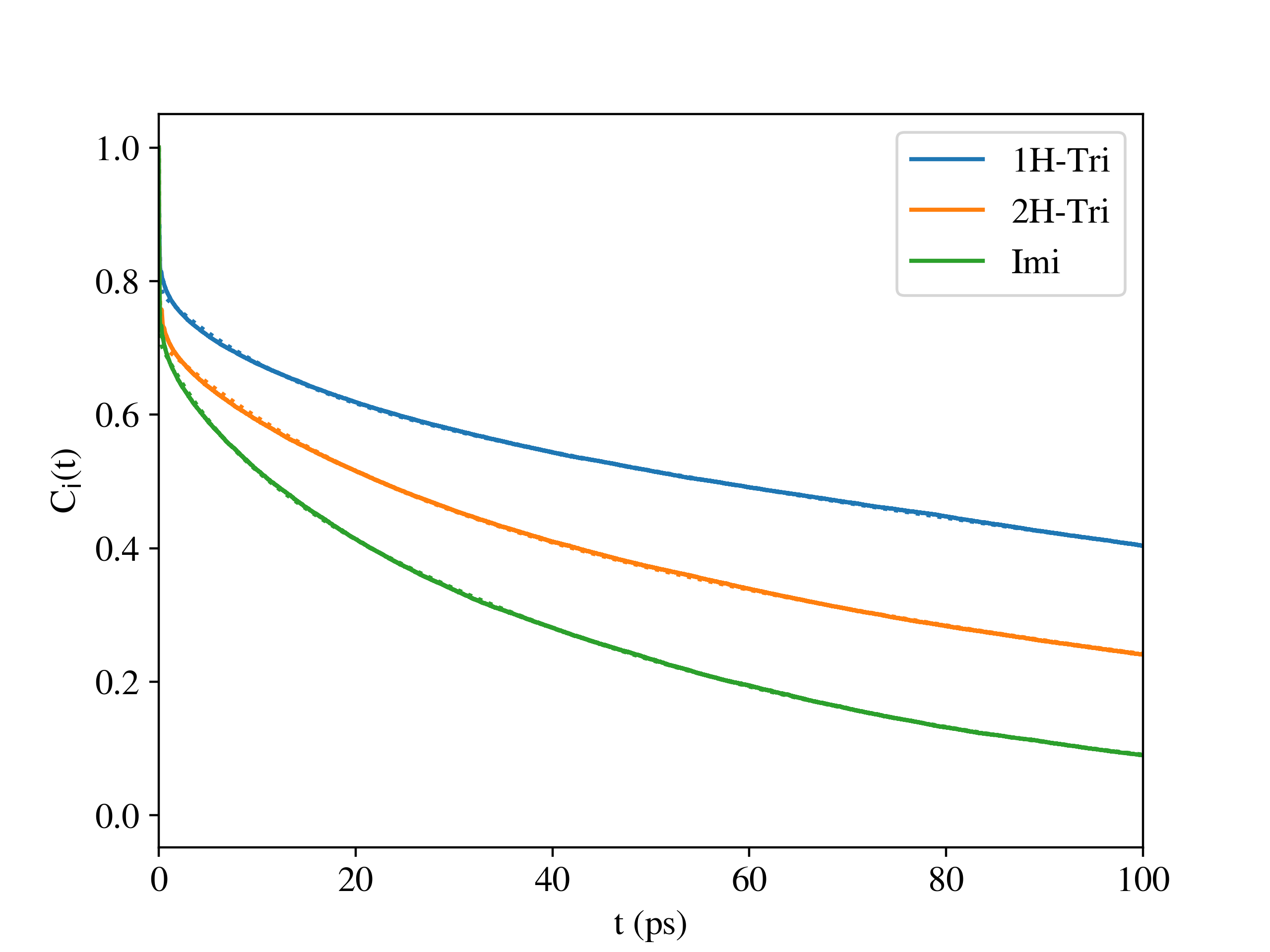}
    \caption{Hydrogen bond population correlation functions for Imi, 1H-Tri, and 2H-Tri. Their corresponding triexponential fits are shown as dotted lines.}
    \label{fig:h_corr_funs}
\end{figure}

To analyze hydrogen bond timescales, we measured hydrogen bond population functions, which give the probability that if a hydrogen bond exists between nitrogen atoms A and B at time 0, it will exist at a later time t. The direction of the hydrogen bond is taken into account here, i.e. a hydrogen bonded donated from A and received at B is distinct from the reverse case where B is the donor and A is the acceptor. Similar to the protonation population function, the resulting correlation functions are fit to a triexponential form. The timescales of the fit are shown in Fig.~\ref{fig:h_corr_funs} and Table~\ref{table:hbondcorr}

\begin{table}[ht]
\setcellgapes{3pt}
\makegapedcells
\centering
\begin{tabularx}{\linewidth}{LRRR}
\toprule[1pt]
System & $\tau_1$ & $\tau_2$ & $\tau_3$\\
\hline
Imidazole (384 K) & 0.0625 & 6.04 & 53.2 \\
1-\textit{H}-1,2,3-triazole (300 K) & 0.0846 & 11.3 & 207 \\
2-\textit{H}-1,2,3-triazole (300 K) & 0.0790 & 19.5 & 128 \\
\bottomrule[1pt]
\end{tabularx}
\caption{Time constants for the hydrogen bond population function in Imi, 1H-Tri, and 2H-Tri}
\label{table:hbondcorr}
\end{table}

\section{Proton Transfer Videos}

These three videos obtained from our simulations show some of the proton transfer mechanisms discussed in the main text. In all the videos, the excess protons are marked in red.

\begin{itemize}
\item \textbf{Imidazole (Imi.mp4)}: Shows the back-and-forth proton transfer (rattling) between two imidazole molecules in a hydrogen-bonded chain.
\item \textbf{1-\textit{H}-1,2,3-triazole(1H-Tri.mp4)}: Shows the proton rattling between three 1H-Tri hydrogen-bonded molecules.
\item \textbf{2-\textit{H}-1,2,3-triazole(2H-Tri.mp4)}: Shows a 1H-Tri molecule (tautomerized from 2H-Tri) adjacent to a triazolium molecule. The 1H-Tri molecule rotates so that its N3 position faces the excess proton, after which it accepts the excess proton to form 1,3-di-\textit{H}-1,2,3-triazolium.
\end{itemize}

\end{document}